\documentstyle[psfig]{mn}

\newif\ifAMStwofonts

\let\tableline=\hline

\def\Romain#1{\expandafter\uppercase\expandafter{\romannumeral #1}}
\def\ion#1#2{#1$\;${\small\rm\Romain#2}\relax}

\ifoldfss

  \newcommand{\bld}[1] {{\bf #1}}

  \ifCUPmtlplainloaded \else
    \NewTextAlphabet{textbfit} {cmbxti10} {}
    \NewTextAlphabet{textbfss} {cmssbx10} {}
    \NewMathAlphabet{mathbfit} {cmbxti10} {} % for math mode
    \NewMathAlphabet{mathbfss} {cmssbx10} {} %  "   "    "
  \fi

  \ifAMStwofonts

    \ifCUPmtlplainloaded \else
      \NewSymbolFont{upmath} {eurm10}
      \NewSymbolFont{AMSa} {msam10}
      \NewMathSymbol{\upi}     {0}{upmath}{19}
      \NewMathSymbol{\umu}     {0}{upmath}{16}
      \NewMathSymbol{\upartial}{0}{upmath}{40}
      \NewMathSymbol{\leqslant}{3}{AMSa}{36}
      \NewMathSymbol{\geqslant}{3}{AMSa}{3E}

      \let\leq=\leqslant \let\le=\leqslant
       \let\ge=\geqslant
    \fi

  \fi

\fi % End of OFSS

\ifnfssone

  \newmathalphabet{\mathit}
  \addtoversion{normal}{\mathit}{cmr}{m}{it}
  \addtoversion{bold}{\mathit}{cmr}{bx}{it}

  \newcommand{\bld}[1] {\mathbf{#1}}

  \newmathalphabet{\mathbfit} % math mode version of \textbfit{..}
  \addtoversion{normal}{\mathbfit}{cmr}{bx}{it}
  \addtoversion{bold}{\mathbfit}{cmr}{bx}{it}

  \newmathalphabet{\mathbfss} % math mode version of \textbfss{..}
  \addtoversion{normal}{\mathbfss}{cmss}{bx}{n}
  \addtoversion{bold}{\mathbfss}{cmss}{bx}{n}

  \ifAMStwofonts

    \ifCUPmtlplainloaded \else

      \UseAMStwoboldmath

      \makeatletter
      \new@mathgroup\upmath@group
      \define@mathgroup\mv@normal\upmath@group{eur}{m}{n}
      \define@mathgroup\mv@bold\upmath@group{eur}{b}{n}
      \edef\UPM{\hexnumber\upmath@group}

      \new@mathgroup\amsa@group
      \define@mathgroup\mv@normal\amsa@group{msa}{m}{n}
      \define@mathgroup\mv@bold\amsa@group{msa}{m}{n}
      \edef\AMSa{\hexnumber\amsa@group}
      \makeatother

      \mathchardef\upi="0\UPM19
      \mathchardef\umu="0\UPM16
      \mathchardef\upartial="0\UPM40
      \mathchardef\leqslant="3\AMSa36
      \mathchardef\geqslant="3\AMSa3E

      \let\leq=\leqslant \let\le=\leqslant
       \let\ge=\geqslant

    \fi
  \fi

\fi % End of NFSS release 1

\ifnfsstwo

  \newcommand{\bld}[1] {\mathbf{#1}}

  \DeclareMathAlphabet{\mathbfit}{OT1}{cmr}{bx}{it}
  \SetMathAlphabet\mathbfit{bold}{OT1}{cmr}{bx}{it}
  \DeclareMathAlphabet{\mathbfss}{OT1}{cmss}{bx}{n}
  \SetMathAlphabet\mathbfss{bold}{OT1}{cmss}{bx}{n}

  \ifAMStwofonts

    \ifCUPmtlplainloaded \else
      \DeclareSymbolFont{UPM}{U}{eur}{m}{n}
      \SetSymbolFont{UPM}{bold}{U}{eur}{b}{n}
      \DeclareSymbolFont{AMSa}{U}{msa}{m}{n}
      \DeclareMathSymbol{\upi}{0}{UPM}{"19}
      \DeclareMathSymbol{\umu}{0}{UPM}{"16}
      \DeclareMathSymbol{\upartial}{0}{UPM}{"40}
      \DeclareMathSymbol{\leqslant}{3}{AMSa}{"36}
      \DeclareMathSymbol{\geqslant}{3}{AMSa}{"3E}

      \let\leq=\leqslant \let\le=\leqslant
       \let\ge=\geqslant

    \fi
  \fi

\fi % End of NFSS release 2

\ifCUPmtlplainloaded \else
  \ifAMStwofonts \else % If no AMS fonts
    \def\upi{\pi}
    \def\umu{\mu}
    \def\upartial{\partial}
  \fi
\fi

\title[Spectral properties of galaxies in the SAPM survey]{Spectral
Analysis of the Stromlo-APM Survey\\ I. Spectral Properties of
Galaxies}

\author[L. Tresse, S. Maddox , J. Loveday and C. 
Singleton]{L.~Tresse,$^{1,2,3}$ 
S.~Maddox,$^{1,5}$ J.~Loveday$^{4}$ and C.~Singleton$^{1}$ \\
$^1$Institute of Astronomy, Cambridge, CB3 0HA, UK \\
$^2$Laboratoire d'Astronomie Spatiale, Traverse du Siphon, B.P.8, 
13376, Marseille Cedex 12, France \\
$^3$Istituto di Radioastronomia del CNR, Via P. Gobetti, 101, 
40129 Bologna, Italia \\
$^4$Department of Astronomy and Astrophysics, University of Chicago, 
5640 S. Ellis Ave, Chicago, IL 60637, USA  \\
$^5$School of Physics and Astronomy, University of Nottingham, Nottingham, 
NG7 2RD, UK}

\pagerange{\pageref{firstpage}--\pageref{lastpage}}
\pubyear{1999}

\begin{document}

\maketitle

\label{firstpage}

\begin{abstract}
We analyze spectral properties of 1671 galaxies from the Stromlo-APM
survey, selected to have $15 \leq b_J \leq 17.15$ and having a mean
redshift $z = 0.05$.  This is a representative local sample of field
galaxies, so the global properties of the galaxy population provide a
comparative point for analysis of more distant surveys.  We measure
H$\alpha$, [\ion{O}{2}]~$\lambda3727$,
[\ion{S}{2}]~$\lambda\lambda$6716, 6731, [\ion{N}{2}]~$\lambda$6583
and [\ion{O}{1}]~$\lambda$6300 equivalent widths and the D$_{4000}$
break index.  The 5 \AA\ resolution spectra use an 8\arcsec\ slit,
which typically covers 40--50\% of the galaxy area. We find no
evidence for systematic trends depending on the fraction of galaxy
covered by the slit, and further analysis suggests that our spectra
are representative of integrated galaxy spectra.

We classify spectra according to their H$\alpha$ emission, which is
closely related to massive star formation. Overall we find $61\%$ of
galaxies are H$\alpha$ emitters with rest-frame equivalent widths
EW(H$\alpha$) $\ga 2$ \AA.  The emission-line galaxy (ELG) fraction is
smaller than seen in the CFRS at $z = 0.2$ \cite{tre96} and is
consistent with a rapid evolution of H$\alpha$ luminosity density
\cite{tre98a}.  The ELG fraction, and EW(H$\alpha$), increase at
fainter absolute magnitudes, smaller projected area and smaller
D$_{4000}$. In the local Universe, faint, small galaxies are dominated
by star formation activity, while bright, large galaxies are more
quiescent. This picture of the local Universe is quite different from
the distant one, where bright galaxies appear to show
rapidly-increasing activity back in time.

We find that the ratio [\ion{N}{2}] $\lambda$6583/H$\alpha$ is
anti-correlated with EW(H$\alpha$), and that the value of 0.5 commonly
used to remove the [\ion{N}{2}] contribution from blended H$\alpha$ +
[\ion{N}{2}] $\lambda\lambda$6548, 6583 applies only for samples with
an EW distribution similar to that seen at low redshift.  We show that
[\ion{O}{2}], [\ion{N}{2}], [\ion{S}{2}] and H$\alpha$ EWs are
correlated, but with large dispersions ($\sim50\%$) due to the
diversity of galaxy contents sampled. Our [\ion{O}{2}]--H$\alpha$
relation is similar to the one derived by Kennicutt \shortcite{ken92},
but is a factor $10\%$ higher at $1\sigma$ significance. We show that
this relation is not valid for distant, strong [\ion{O}{2}] emitters
with blue colors, which are more numerous than locally. This relation
would overestimate the individual star formation rate (SFR) by $\sim50\%$
for these kind of galaxies.  We find that on average luminous blue ELG
are likely to be enhanced in nitrogen abundance.  This suggests that
in faint, low-mass, late-type ELG, nitrogen is a primary element,
whereas in more bright, more massive galaxies nitrogen it comes from a
secondary source. We also find $4\%$ of early-type galaxies
which show star formation activity; this fraction seems to increase at
higher redshifts.
\end{abstract}


\section{Introduction}

Studies of spectral properties within well-defined surveys are crucial
to determine the evolutionary properties of galaxies.  Indeed the
continuum of spectra provides information on the global stellar
content, while the lines are powerful diagnostics of the star
formation rate. To understand the evolution of galaxies from today to
earlier epochs, one major task is to carefully compare different
surveys. Each survey has its own galaxy selection and line detection
levels, and these may vary even within one survey if it samples a
large range of redshifts.

The main aim of this paper is to measure the ave\-rage spectral
properties of a representative local $b_J$-selected galaxy population,
which can be used as a reference point for more distant galaxy
samples.  We analyse spectra from the Stromlo-APM redshift survey,
which uniformly samples a large volume and gives a representative
population of local galaxies.  The measured spectral properties
include H$\alpha$, [\ion{O}{2}] $\lambda$3727, [\ion{N}{2}]
$\lambda$6583, [\ion{S}{2}] $\lambda\lambda$6716, 6731, [\ion{O}{1}]
$\lambda$6300 xand the 4000 \AA\ break. Most previous studies have
concentrated on [\ion{O}{2}], since it is the line commonly observed
at larger redshifts.  Combined with other lines, and photometric
properties, a deeper knowledge of galaxy content and star formation
rate evolution can be reached.

We wish to stress that our aim is to quantify the {\it global}
spectral properties of the galaxy population. Measurements for
individual galaxies may be affected by the limited areal coverage of
the spectral slit, but the overall average measurements should be
accurate as we demonstrate through this paper.  Our measurements of
spectral features are flux ratios (EWs and the 4000 \AA\ Balmer break
index), and thus they are insensitive of uncertainties in the flux
calibration.

This paper is organized as follows. Section 2 describes the
spectroscopic database, our measurements, and the slit effects.  In
Section 3 we classify spectra according to their H$\alpha$ emission,
and investigate how the fraction of H$\alpha$ emitters (EW(H$\alpha$)
$\ga 2$ \AA) depends on other galaxy properties.  In Section 4 we
compare this to other surveys.  Sections 5, 6 and 7 present the
properties of the H$\alpha$-emitter population; we also examine the
repercussions for star formation rate measurements.  Section 8
discusses those galaxies which exhibit [\ion{O}{2}] in the absence of
H$\alpha$ emission.  We conclude in Section 9.

\section{Data}

\subsection{The spectroscopic survey} 

Our sample of galaxies is taken from the Stromlo-APM (SAPM) redshift
survey which covers 4300 sq-deg of the south galactic cap and consists
of 1797 galaxies brighter than $b_J = 17.15$ mag.  The galaxies all
have redshifts $z < 0.145$, and the mean is $\langle z \rangle =
0.051$.  A detailed description of the spectroscopic observations and
the redshift catalog is published by Loveday et
al. \shortcite{lov96a}.  In this paragraph, we briefly recall some
important points.

The SAPM galaxies were randomly selected at a rate of 1 in 20 from
Automated Plate Measurement (APM) machine scans of UK-Schmidt (UKS)
plates (see Loveday 1996; Maddox et al. 1990).  Most of the galaxies
were given a morphological class from visual inspection of the UKS
plates, and we will investigate the link between spectral properties
and morphology in a future paper.  For the faintest galaxies
morphological classification is difficult, and many galaxies were not
given a classification, leading to incompleteness in the morphological
subsamples \cite{zuc94}.  Our present study is not affected by this
problem, since we study all galaxy spectra, whether or not they have
been assigned to a specific morphological class.

The galaxy images that overlap other images were excluded from the
redshift catalog (214 out of 2011) to guarantee reliable redshift and
magnitude measurements.  Most of these galaxy images overlap with star
images ($70.5\%$ of the overlaps), the rest with other galaxy images.
The galaxy--star overlapping images are a random sub-sample of the
galaxies, and thus their rejection does not affect our analysis.
Rejecting galaxy--galaxy overlapping images may introduce a bias
against physically merging systems.  However galaxy--galaxy overlaps
represent only $3\%$ of the total galaxy ima\-ges, hence the fraction of
genuine merging systems should be smaller than this.

Of the 1797 galaxies originally published in the redshift survey, 54
have a redshift taken from the literature (of which 28 are at $b_J <
15$), and for 7 we could not retrieve the spectra because they were
not observed with the Dual-Beam Spectrograph (DBS) of the ANU 2.3-m
telescope at Siding Spring.  Amongst the 1736 spectra we had in hand,
we excluded 7 blueshifted spectra, 3 with $c z < 1000$~km~s$^{-1}$,
and 2 with too low signal-to-noise.  Following Loveday et
al. \shortcite{lov92}, we also applied a bright apparent magnitude cut
to the sample to reject galaxies which have saturated ima\-ges on
photographic plates.  Thus, in this paper, we studied 1671 spectra
with $15 \leq b_J \leq 17.15$, $c z > 1000$ km s$^{-1}$, which
constitute a representative sample of nearby gala\-xies.  It has been
argued that the SAPM survey is biased by photometric calibration
errors, but in addition to the original CCD photometry (Maddox et
al. 1990b; Loveday 1996), new CCD photometry has recently been used to
check the SAPM calibration, and this shows no evidence of significant
photometric errors \cite{lov99}.  In any case our measurements of EW
and 4000 \AA\ Balmer break index are independent of photometric
calibration and so our results are not affected by uncertainties in
calibration.

\begin{figure}
\psfig{figure=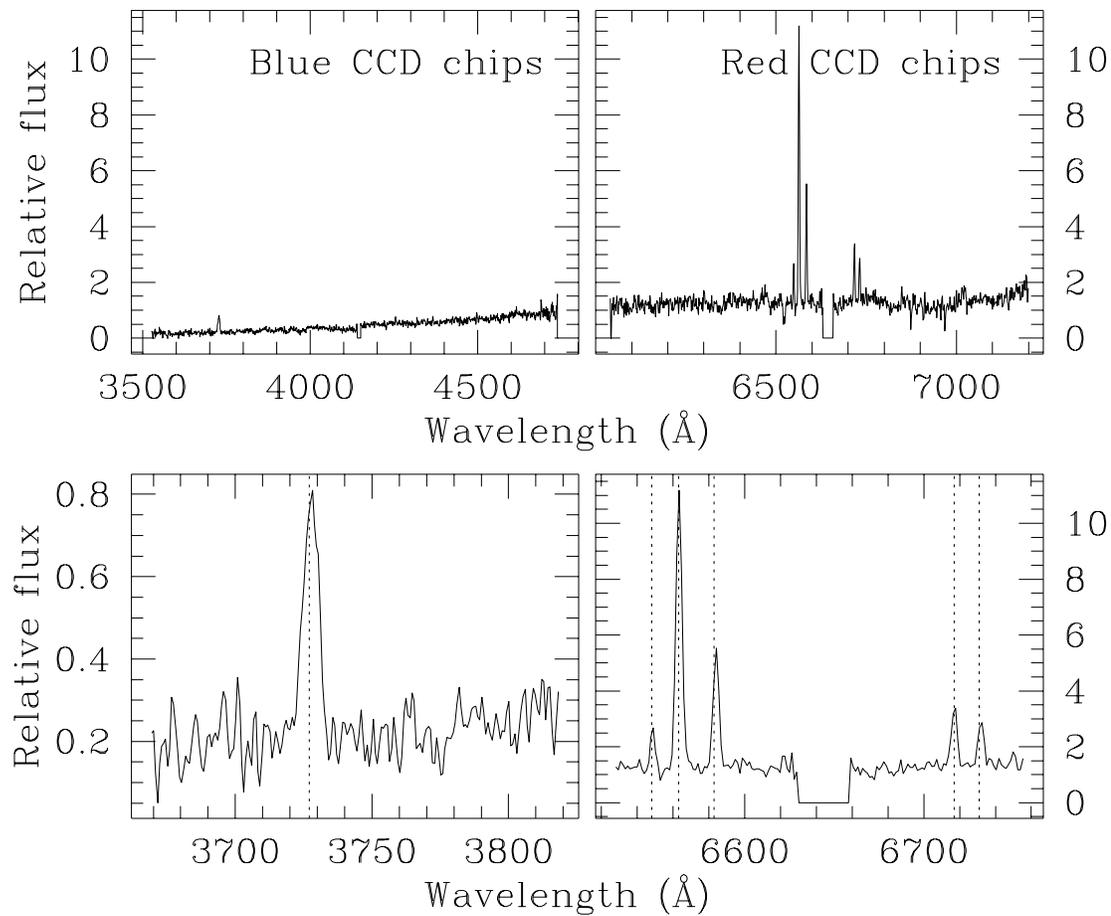,height=15cm} 
\caption[plot00]{ Top panels: The complete rest-frame wavelength range
spanned by one SAPM spectrum at $z = 0.0534$. One can see the small
gaps in continuum at (observed frame) $\sim$[4360--4370]\AA, and
$\sim$[7000--7020]\AA\ (see Section 2.1).  Bottom-left panel: Zoomed
view of [\ion{O}{2}] $\lambda$3727. Bottom-right panel: Zoomed view of
[\ion{N}{2}] $\lambda$6548, H$\alpha$, [\ion{N}{2}] $\lambda$6583, 
[\ion{S}{2}] $\lambda\lambda$6716, 6731 respectively.
\label{plot00}}
\end{figure}

The DBS divides the light into blue and red beams. The wavelength
range is 3700--5000 \AA\ in the blue, and 6300--7600~\AA\ in the red,
both with a dispersion of $\sim1$ \AA\ per pixel.  Each beam is
fibre-coupled to two CCD chips; this leaves small gaps in the
wavelength coverage between 4360--4370 \AA\ in the blue, and
7000--7020 \AA\ in the red (see Fig.~\ref{plot00}).  The mean
integration time for the 1671 galaxies is 470s, of which 795 have a
600s exposure time. Each integration was stopped when spectral
features could be identified at the level required to measure a
redshift. As a consequence the variation in signal-to-noise as a
function of apparent magni\-tude is reduced. For instance the
signal-to-noise of EW(H$\alpha$) varies by only a factor $\sim2$ over
our observed apparent magni\-tude range, whilst a fixed exposure time
would have given a factor $\sim10$.  The width of $\sim8$\arcsec\ slit
leads to a spectral resolution of about FWHM $=5$ \AA.

We calibrated spectra in {\it relative} flux using spectrophotometric
standard stars.  Since standard stars were not observed systematically
each night, we cannot have reliable absolute flux calibration.
Consequently we built a mean sensitivity function from the several
standard stars observed in each run, and applied it to all objects
observed during the corresponding run. Hence the zero-point of flux
calibration cannot be properly recovered, and the calibration can only
be relative.  That is equivalent to correcting for the instrumental
response function and atmospheric extinction, and to applying a gross
zero-point for flux calibration.

By comparing continuum levels across the two blue CCDs we estimate
that the relative calibration is good to $\sim1\%$; the accuracy is
the same for the two red CCDs.  Comparing the continuum in the red and
blue spectra suggests that their relative calibration is consistent to
$\sim5\%$, though occasionally the discrepancy appears to be as much
as $20\%$.  However, the large gap between the red and blue spectra
makes comparison of continuum levels liable to large uncertainties.
Since we concentrate our analyses on EW measurements, calibration
errors at this level are not important.

\subsection{Emission-line and D$_{4000}$ measurements}

The prominent spectral features include in the blue range [\ion{O}{2}]
$\lambda$3727, Ca$\;$H\&K $\lambda\lambda$3933, 3969, the 4000 \AA\
Balmer break and for a few galaxies H$\beta$. In the red range we
observe H$\alpha$, [\ion{N}{2}] $\lambda\lambda$6548, 6583,
[\ion{S}{2}] $\lambda\lambda$6716, 6731, and for some strong
emission-line spectra [\ion{O}{1}] $\lambda$6300.  Unfortunately
[\ion{O}{3}] $\lambda\lambda$4959, 5007, and usually H$\beta$, are in
the gap between the blue and red parts of the spectra (from $\sim5000$
to $\sim6300$~\AA).

We measured the integrated relative fluxes, $F$, and rest-frame
equivalent widths, EW, of the emission lines using the package {\sc
splot} under {\sc iraf/cl}, interactively marking two endpoints around
the line to be measured.  This method allows the measurement of lines
with asymmetric shapes (i.e. with deviations from Gaussian profiles).
In most cases as in Fig.~\ref{plot00} the H$\alpha$ and [\ion{N}{2}]
lines could be measured separately. When these lines are blended, we
used the Gaussian de\-blending program. Note that in these later cases
lines are only partly blended.  The interactive method allows us to
control by eye the level of the continuum taking into account defects
that may be present around the line measured.  It does not have the
objectivity of automatic measurements, but it does allow us to obtain
reliable accurate measurements.

We estimated $1\sigma$  standard deviations as follows: 
$$
\sigma_{F}  =  \sigma_{c} D \sqrt{2 N_{pix} + {\rm EW} / D } $$ 
for the flux, and 
$$\sigma_{\rm EW} = \frac{\rm EW}{\rm F} \sigma_{c} D \sqrt{ {\rm EW}
/ D + 2 N_{pix} + ({\rm EW} / D)^2 / N_{pix} } $$ for the EW, where
$D$ is the spectral dispersion in \AA\ per pixel, $\sigma_{c}$ is the
mean standard deviation per pixel of the continuum on each side of the
line, and $N_{pix}$ is the number of pixels covered by the line.  Note
that these are not exactly the formal statistical $1\sigma$ errors of
$F$ or EW, because we estimate the signal-to-noise for each pixel by
scaling the continuum variance according to Poisson statistics.  Our
approximation slightly overestimates the errors, but in our analysis
we are mainly interested in the consistency of the estimation of
errors.

\begin{figure}
\psfig{figure=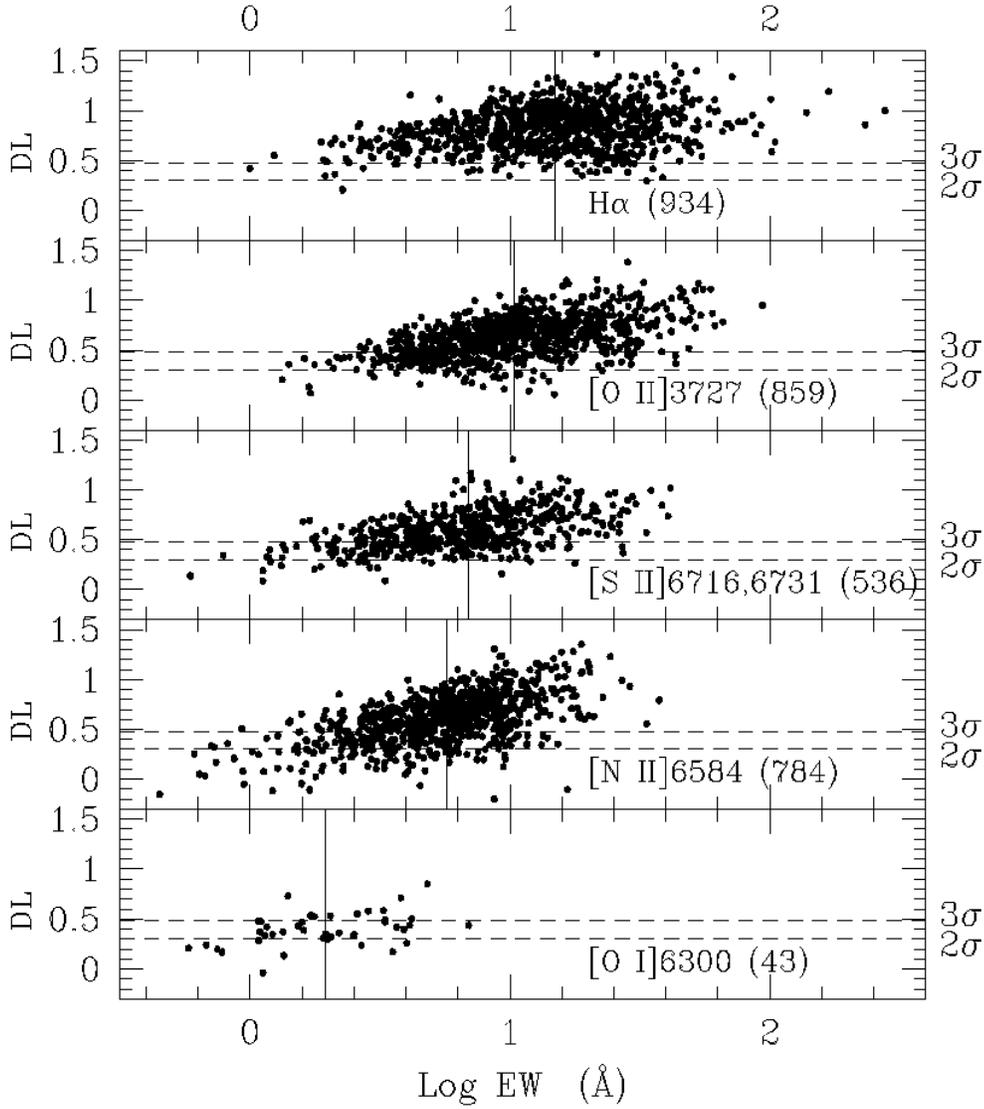,height=15cm} 
\caption[plot2]{ Top panel: Log of the detection level of
EW(H$\alpha$), i.e. DL $=$ Log~(EW/$\sigma$(EW)), versus Log
EW(H$\alpha$). Lower panels: Same but for [\ion{O}{2}] $\lambda$3727,
for [\ion{S}{2}] $\lambda\lambda$6716, 6731, for [\ion{N}{2}]
$\lambda$6584, and for [\ion{O}{1}] $\lambda$6300. Vertical solid lines are the
respective EW medians. The number of plotted data and the 2,
$3\sigma$ detection levels (dashed lines) are indicated in each panel.
\label{plot2}}
\end{figure}

In our study, we use mainly the EW, since we have only relative
fluxes.  Whenever we could see a significant H$\alpha$ emission line
by eye, we measured the EW.  Figure~\ref{plot2} shows that our EW
limit is approximately at a $3\sigma$ confidence level.  For the
signal-to-noise of our spectra this corresponds to measurements of
EW(H$\alpha) \ga 2$\AA, and the average detection level is
$5.6\sigma$.  Then [\ion{N}{2}] $\lambda$6583 was measured;
[\ion{N}{2}] $\lambda$6548 is 3 times less intense and so is usually
too faint to be detected or has often a too low signal-to-noise to be
detected in the same homegeous manner as its counterpart.  As seen in
Figure~\ref{plot2} the [\ion{N}{2}] measurements include lines
detected at less than $2\sigma$, because it was simple to attempt a
measurement whenever we measured H$\alpha$.  Detection at the
$3\sigma$ confidence level corresponds to EW([\ion{N}{2}]
$\lambda$6583) $\ga 2$ \AA, similar to the H$\alpha$ limit.  We
measured both the [\ion{S}{2}] $\lambda$6716 and [\ion{S}{2}]
$\lambda$6731 lines, and added their EW to give [\ion{S}{2}]
$\lambda\lambda$6716, 6731 as plotted in Figure~\ref{plot2}. If only
one [\ion{S}{2}] line could be seen above $2\sigma$, the EW of the
non-detected line was set to zero. Again the $3\sigma$ detection level
is $\sim2$ \AA.  Although [\ion{O}{1}] $\lambda$6300 is rarely
detectable, we were able to measure it in a few spectra. The $3\sigma$
detection level is also at $\sim2$ \AA. 
The [\ion{O}{2}] measurements typically have confidence levels $\ga
2\sigma$. Its $3\sigma$ detection level is 3 \AA, slightly higher than
the other lines, because the blue continuum is less intense than the
red, and so is more noisy.  All of the lines in the red part of the
spectrum have a $3\sigma$ detection limit of $\sim2$ \AA\ because the
signal-to-noise of the red continuum is about the same for each of
these lines. This shows that our measurements were made in a
consistent way.

As well as the emission lines, we also measured the 4000~\AA\ Balmer
break.  This spectral discontinuity is due to the opacity produced by
the presence of a large number of spectral lines of ionized metal, in
a narrow wavelength region. Its amplitude depends on the metallicity,
and thus on the stellar temperature and age \cite{bru83}. Early-type
galaxies usually have larger amplitudes than late-type galaxies.  We
estimated the break as follows:
$$D_{4000} = f(4200 - 4000) / f(4000 - 3800), $$ where $f$ is the mean
integrated flux within the rest-frame wavelength range mentioned.  We
rejected pixels having fluxes more than $2\sigma$ from the mean flux;
this ensured that emission or absorption lines and bad pixels did not
bias the measurements.

\subsection{Slit effects}

The galaxy spectra were taken using a slit 8\arcsec\ wide and
7\arcmin\ long.  The slit was always positioned to cross the central
region of the galaxies, thus the light collected for each galaxy
originates from the core and from a fraction of the outer part of the
galaxies.  In this section, we assess how close our spectra are to
fully integrated galaxy spectra.

\begin{figure}
\psfig{figure=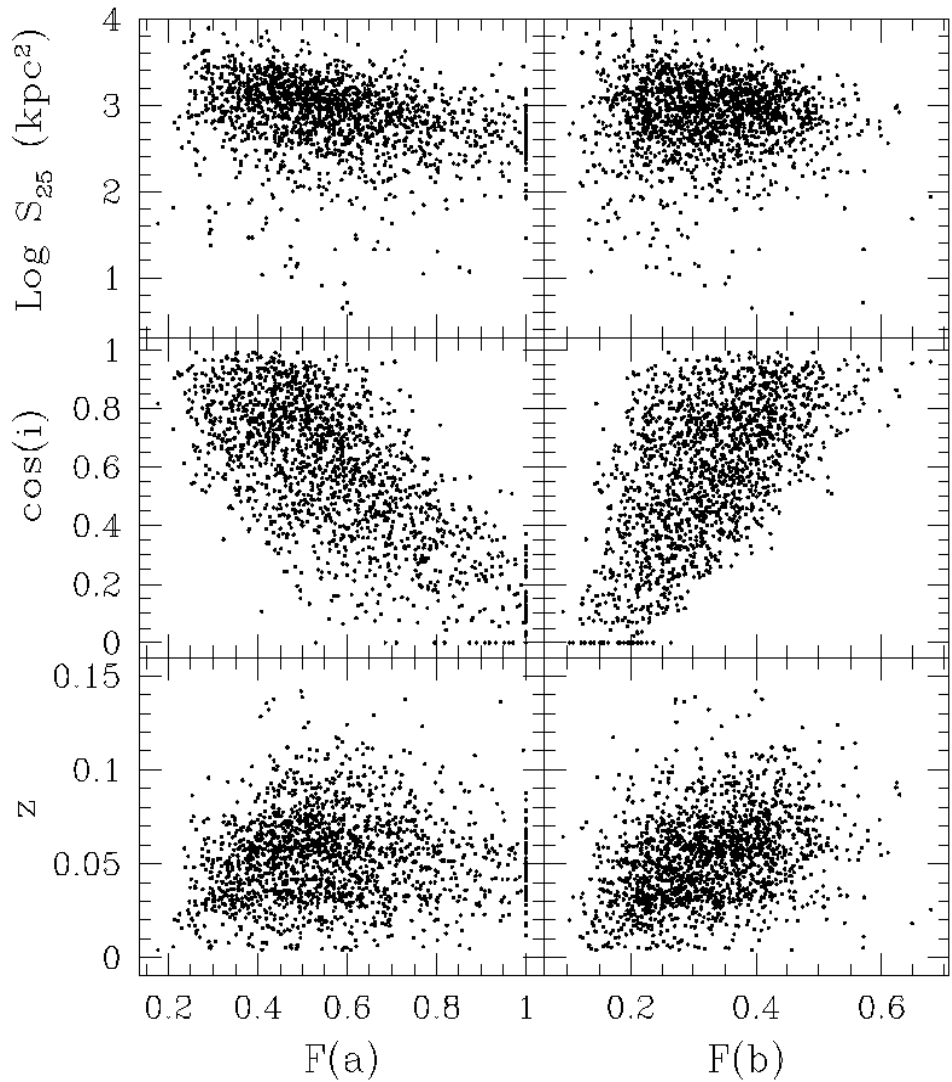,height=15cm}
\caption[plot1]{ For the whole sample (1671), fractions of the galaxy
image covered by the 8\arcsec\ wide slit if the slit was positioned
along the major axis, $F(a)$, or the minor axis, $F(b)$, as a function
of the physical size at 25 mag arcsec$^{-2}$, $S_{25}$ (upper panels),
the cosine inclination angle, cos($i$) (middle panels), and the
redshift, $z$ (lower panels).
\label{plot1}}
\end{figure}

We estimated the area of each galaxy image using the measurements of
the major and minor axes, $a$ and $b$, at a $b_J$ surface brightness
level of 25 mag arcsec$^{-2}$ (see Maddox et al. 1990).  The major
axis $a$ varies between 15\arcsec\ and 153\arcsec\ with $\langle a
\rangle = 36$\arcsec, and so is always shorter than the length of the
slit.  The minor axis $b$ is between 6\arcsec\ and 67\arcsec\ with
$\langle b \rangle = 21$\arcsec, and so is usually larger than the
slit width.  The orientation of the slit was not systematically
recorded, therefore we do not know the angle of the slit relative to
the major axis of the galaxies.  However, we can test the two possible
extreme cases of overlaying each galaxy image with an 8\arcsec\ slit
positioned either along the major axis, or along the minor axis.  For
the first case, we approximated the fraction of the galaxy image
covered by the slit,
$$ F(a) \approx min \left( \frac{ 8\ a\ } {0.25\ \pi a b } \ ,\
1\right) $$ as a function of $S_{25}$, defined as the rest-frame
projected area\footnote{We assume $H_0 = 50$ km s$^{-1}$ Mpc$^{-1}$
and $q_0 = 0.5$ throughout this paper.} brighter than $b_J =$ 25 mag
arcsec$^{-2}$ in kpc$^2$, and of the galaxy inclination $i$, defined
by $\sin^2(i) = [1 - (b/a)^2]/0.96$ \cite{hub26} (Fig.~\ref{plot1}).
For the second possibility we calculated $F(b)$, using the same
approximation with $8 a$ replaced by $8 b$.  The true configuration
must lie between these two possible extremes.  With an 8\arcsec\ wide
slit, this fraction depends mainly on the galaxy inclination, $i$, rather
than on $S_{25}$ (Table~\ref{table1}).  As seen in Figure~\ref{plot1},
our typical spectrum samples the light from $\sim45\%$ of the
rest-frame projected area of the galaxy. 
The Figure and Table show also that $F(a)$ and $F(b)$ are respectively
2\% and 7\% larger at $z > 0.05$ than at $z < 0.05$. Thus the average
fraction of galaxy projected area covered by our slit does not depend
strongly on redshift; the high-$z$ galaxies have only 5\% more
coverage than the low-$z$ ones.

\begin{figure}
\psfig{figure=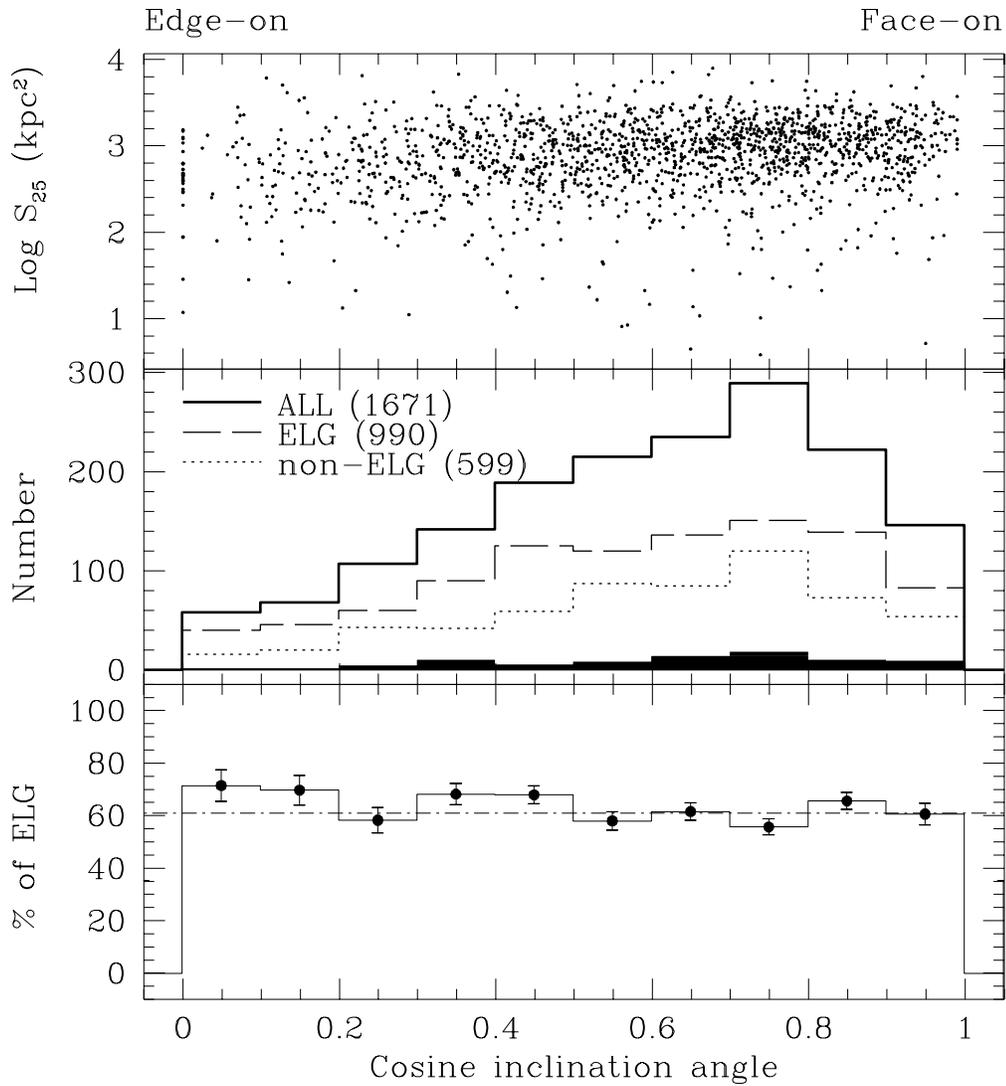,height=15cm} 
\caption[plot1b]{ Top panel: Physical size at 25 mag arcsec$^{-2}$,
$S_{25}$, as a function of the cosine inclination angle,
cos($i$). Middle panel: Distributions of cos($i$), the shaded
histogram is for the 82 unclassified galaxies (see Sect.~3.1). Bottom panel:
Fractions of ELG as a function cos($i$), the mean is $61\%$
(dashed-dotted line).
\label{plot1b}}
\end{figure}

Figure~\ref{plot1b} shows that $\cos(i)$ and $S_{25}$ are barely
correlated, which demonstrates that the SAPM survey is representative
of the different kinds of nearby galaxies at $15 \leq b_J \leq 17.15$.
A galaxy sample with random inclinations would have a uniform
distribution as a function of $\cos(i)$, but our sample is biased
towards galaxies with low inclination angles, as expected in
magnitude-limited surveys \cite{mai95}.  This is because the more
inclined a galaxy is, the more its intrinsic galactic absorption
reduces the light along the line of sight. Thus with the same absolute
luminosity and distance, an inclined galaxy is likely to be fainter
than a face-on galaxy. This effect will be particularly strong in
blue-selected surveys where dust absorption is large.

\section{ELG fraction} 

\subsection{Classification of spectra} 

We classified spectra according to the H$\alpha$ recombination line;
when it is detected in emission as an emission-line galaxy (ELG),
otherwise as a non emission-line galaxy (non-ELG).  The level of
detection of H$\alpha$ depends on the intensity of the line, on its
equivalent width, and on the signal-to-noise of the continuum.  Thus
when comparing the ELG fraction to other surveys, one has to bear in
mind the limits of detection of emission lines.  Another point is that
H$\alpha$ may suffer from stellar absorption, reducing the EW of
emission by as much as 5 \AA\ (see Kennicutt 1992), especially in the
case of early-type galaxies. Weak H$\alpha$ lines ($\ll 10$ \AA) may
not be detected, which would lead to a non-ELG classification.
 
An ELG classification scheme based on H$\alpha$ has the following
advantages: (a) Comparison of H$\alpha$ to [\ion{N}{2}], [\ion{S}{2}]
and [\ion{O}{1}] allows discrimination between AGN and
\hbox{\ion{H}{2}} galaxies \cite{vei87}.  (b) It is a tracer of recent
star formation for most galaxies \cite{ken92}.  (c) Of the Balmer
lines, it is the most directly proportional to the ioni\-zing stellar UV
flux at $\lambda < 912$ \AA\ (see Osterbrock 1989 for a review;
Schaerer \& de Koter 1997 for recent models), because the weaker
Balmer lines are much more affected by the equivalent absorption lines
produced in stellar atmospheres.  (d) It does not depend strongly on
the metallicity, which is not the case for the other commonly observed
optical lines such as [\ion{N}{2}], [\ion{S}{2}], [\ion{O}{2}] and
[\ion{O}{3}].  Since these latter are forbidden lines, their
presence depends also on the density of the gas. They have higher
ionizing potential than Balmer lines, thus depend also on the hardness
of the ionizing stellar spectra.  (e) It is less affected by dust than
any other lines at shorter wavelengths.  Consequently a classification
based on H$\alpha$ has the advantage that it is more or less
comparable at any redshift with less dependence on chemical evolution
than other spectral features.

When H$\alpha$ falls in the wavelength coverage gap $\sim$7000--7020
\AA\ (see Section 2.1), we cannot see whether this line is in emission
or not. To avoid introducing any bias, we systematically did not
classify these spectra, even if other lines strongly suggest that
H$\alpha$ should indeed be in emission or absorption.  There are 82
spectra in this category, which represent $5\%$ of the sample of 1671
spectra.  990 ($59\%$) spectra have been classified as H$\alpha$
emitters, i.e. as ELG, and 599 ($36\%$) as non-H$\alpha$ emitters,
i.e. as non-ELG.  If we assume that the 82 unclassified spectra are
distributed like the classified ones, this give $62\%$ of ELG with
EW(H$\alpha$) $\ga 2$ \AA, and $38\%$ of non-ELG.  However the red gap
location means that these unclassified galaxies are mainly at $z \sim
0.06$ (see Fig.~\ref{plot6}), and thus they are mainly bright
(M($b_J$)$<-$21 in Fig.~\ref{plot7}) where, as we will see, the fraction
of ELG is smaller than at fainter magnitudes.  When we do attempt to
classify the 82 spectra by taking into account the presence of other
spectral lines, we find the percentage of ELG and non-ELG is $61\%$
and $39\%$. This is essentially the same as assuming that the 82
spectra are distributed like the classified ones, and so we conclude
that any bias due to H$\alpha$ falling in the wavelength coverage gap
is insignificant.

\subsection{ELG fraction and slit effects} 

Since the SAPM spectra are not integrated over the whole galaxy, the
fraction of ELG with EW(H$\alpha$) $\ga 2$ \AA\ may be slightly higher
than estimated in Section 3.1.  Indeed, one strong \hbox{\ion{H}{2}}
region outside the central galaxy region can change an absorption-like
spectrum into an ELG. Thus if the 8\arcsec\ wide slit did not cover
this \hbox{\ion{H}{2}} region, a galaxy would be classified as non-ELG
rather than ELG.

However, since the spectral slit covers on average 45--50\% of the
galaxy, and since \hbox{\ion{H}{2}} regions are mainly distributed
along the spiral arms, our fraction of ELG should remain approximately
correct. Even though the distribution of \hbox{\ion{H}{2}} regions
tends to be more centrally concentrated in bulge dominated systems,
the luminous \hbox{\ion{H}{2}} regions tend to be found at larger
radius in late-type spirals (see for instan\-ce Hodge \& Kennicutt
1983). No bias should arise for galaxies with central H$\alpha$
emission. The major caveat might come from irregular systems where
star-forming regions can be found anywhere.  In the SAPM sample,
galaxies classified as irregular represent only 5\%. The
morphologically unclassified galaxies (21\%; see Table~1 in Loveday,
Tresse \& Maddox 1999) typically have small images with no easily
identifiable morphological features. They are likely to be early
Hubble-type galaxies, or spirals with weak spiral arms, or compact
galaxies, and are very unlikely to be irregulars. Those galaxies with
only one dominant \hbox{\ion{H}{2}} region in the disk (perhaps due to
an interaction with another galaxy for instance), should represent a
very small fraction. In these cases, there was $\sim50\%$ chance for
having observed the \hbox{\ion{H}{2}} region with an 8\arcsec\ slit.
Moreover, we note that 50 spectra were re-observed because there were
no obvious spectral features to identify accurately a redshift; these
re-observed spectra still remain non-ELG, even though the slit
orientation changed.  Moreover in Section~3.4, we find that very few
non-ELG (11) have a 4000 \AA\ break consistent with what is expected
for blue galaxies, hence it tells us that our fraction of ELG should
be correct to 1\%.

\begin{figure}
\psfig{figure=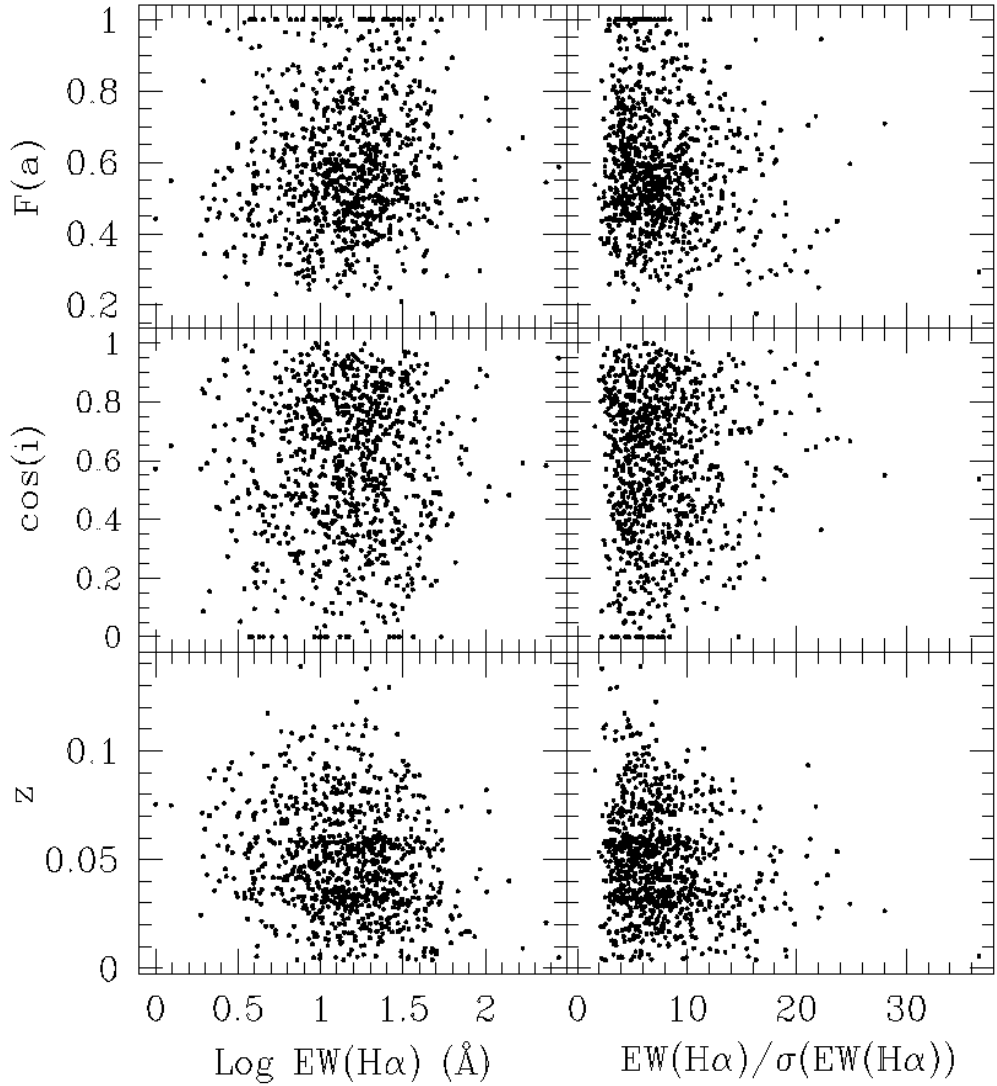,height=15cm} 
\caption[plot1c]{ Log EW(H$\alpha$), and detection level of
EW(H$\alpha$) as a function of the fraction of the galaxy image
covered if the slit was positioned along the major axis, $F(a)$ (upper
panels), the cosine inclination angle, cos($i$) (middle panels), and
the redshift, $z$ (lower panels).
\label{plot1c}}
\end{figure}

Figure~\ref{plot1c} shows that neither EW(H$\alpha$) $\ga 2$ \AA\ nor
the signal-to-noise of the measurements depend globally on the galaxy
inclination angle, $i$, or the fraction of area covered by the
8\arcsec\ slit.  Consequently we find that the SAPM sample shows no
trend to detect more ELG than non-ELG as a function of inclination
angles $i$ (Fig.~\ref{plot1b}).  The means of the axial ratios,
($b/a$), are similar for ELG and non-ELG, respectively 0.60 (or $i =
55\degr$) and 0.63 (or $i = 52\degr$).

\subsection{Correlation of the ELG fraction with galaxy properties} 

The average parameters ($b_J$, $z$, $kz$, $M(b_J)$, $\mu_{25}$,
$S_{25}$) for the SAPM galaxies are summarized in Table~\ref{table2}.
The fraction of ELG does not globally depend on $b_J$
(Fig.~\ref{plot5}), showing that the EW is not dependent on apparent
magnitude. This is because the integration times were adjusted for
each galaxy (see Section 2.1).  Figure~\ref{plot6} shows that the ELG
fraction does not systematically depend on redshift, but there is a
slight excess at $z < 0.014$; we discuss this later.

The ELG fraction depends strongly on the absolute magnitudes
(Fig.~\ref{plot7}); from $M(b_J) = -23$ to $-18$ mag, it increases by
a factor of 2.  The fraction also depends on the physical size of
galaxies (Fig.~\ref{plot8}); from $S_{25} = 10 000$ to $100$~kpc$^2$,
it increases by a factor of 4.  ELG are more common than non-ELG in
intrinsically faint galaxies, and small systems.  ELG are on average
$0.43\pm0.06$ mag fainter than non-ELG.  If we exclude galaxies at $z
< 0.014$ (see below), ELG are $0.33\pm0.05$ mag fainter than non-ELG.
The apparent flattening of these trends at $M(b_J) > -18 $ and at
$S_{25} < 40$ is probably due to the combination of the small number
of galaxies in these bins, and the fractions approaching $100\%$.

The dependence on rest-frame surface brightness avera\-ged over the area
brighter than $b_J = 25$ mag arcsec$^{-2}$, $\mu_{25}$, is more
complex.  The fraction of ELG decreases by a factor 2 from $\mu_{25} =
21.9$ to 23.1 mag arcsec$^{-2}$, and increases by the same factor
towards fainter values (Fig.~\ref{plot9}).  Galaxies with $\mu_{25} >
24$ are at $z < 0.014$.  ELG are more common for the highest and
lowest surface brightness galaxies than non-ELG.  This means that
faint, small systems which contribute most to the ELG sample may be
either high-surface brightness, compact galaxies, or low-surface
brightness dwarfs.

Going back to the ELG fraction as a function of $z$ we can now
understand the high fraction at $z < 0.014$.  The SAPM sample is
magnitude-limited with both high and low magnitude limits, and so at
$z < 0.014$ we observe only small, $M(b_J) > -18$ galaxies.  From
Figure~\ref{plot7}, we see that most galaxies this faint are ELG, so
we expect to see a high ELG fraction at these low redshifts.

To summarize, ELG galaxies are found more frequently at fainter
intrinsic magnitudes, and in smaller systems than non-ELG.  This
agrees with the general picture where dwarf and compact galaxies are
more actively forming stars than their bright counterparts in the
local Universe. This is a reflection of the difference between ELG and
non-ELG luminosity functions \cite{lov99b}.  It is also consistent
with the rapidly evolving population of small galaxies up to $z \sim
1$ \cite{bri98}. The overall blue luminosity density is dominated by
bright galaxies, which formed most of their stars much earlier than
dwarfs, which are actively forming stars today.
 
\subsection{Correlation of the ELG fraction with D$_{4000}$}

We estimated the 4000 \AA\ break using the D$_{4000}$ index as defined
in Section 2.2. The measured values of D$_{4000}$ range from 0.9 to
2.5, with a mean of 1.6.  These values are consistent with the limits
obtained with spectral evolution of stellar populations using
isochrone synthesis. For instance in Fig.~13b (D$_{4000}$ versus age,
assuming solar metallicity) of Bruzual \& Charlot \shortcite{bru93}, 
this index never reaches values above 1.5 if a constant star formation
rate is adopted, while a $10^{8.5}$ year old instantaneous-burst or a
$10^{9.5}$ year old 1Gyr-burst reproduce these high values.  Thus half
of the local population is dominated by star bursting of at least
$10^{8.5}$ year old. The other half is dominated by younger star
formation whatever the star formation rates adopted.

The D$_{4000}$ distribution for ELG is clearly different from the
non-ELG distribution (Fig.~\ref{plot4}), the average D$_{4000}$ for
ELG is $31\%$ smaller than for non-ELG (see Table~\ref{table2}).
However, there is an overlap between the two distributions, and there
is no clear separation into distinct populations.  The presence of
H$\alpha$ in emission is correlated with a low D$_{4000}$, which is
expected since both are indicators of massive star formation.  It
strengthens the link between our H$\alpha$ ELG classification and star
formation activity.  In general, the smaller D$_{4000}$, the bluer the
galaxy color; as expected our ELG have bluer colors on average than
the non-ELG.

We note that $1\%$ (12) of ELG galaxies have $D_{4000} > 2$, the
value characteristic of elliptical galaxies. They are bright galaxies
($M(b_J) \leq -21.6$), and have EW(H$\alpha$) $\simeq$
EW([\ion{O}{2}]) $\simeq 8$ \AA. In these systems, stellar absorption
at H$\alpha$ must be significant, and thus their true EW(H$\alpha$)
must be higher, as we expect from the relation between [\ion{O}{2}]
and H$\alpha$ (see Section 7.1).  These galaxies have a significant
old stellar population, and are undergoing current star formation;
most of them have a spiral morphology. 

Also $2\%$ (11) of non-ELG have $D_{4000} < 1.2$, the value typical of
very blue galaxies. They span the whole range of luminosities, they
have no [\ion{O}{2}] line. Five are spirals, two are ellipticals, one 
is irregular and three are not morphologically classified.  For some
of them, this could be due to slit effects as discussed in Section~3.2. 

Figure \ref{plot4bis} shows that nearly all galaxies fainter than
$M(b_J) = -19$ have $D_{4000} < 1.7$. Most dwarf galaxies tend to have
rather young stellar populations, consistent with the high fraction of
ELG that we find at faint magnitudes.

\section{Comparison of the ELG fraction with other surveys} 

Comparing average properties from different surveys is never a
straightforward task.  Different detection limits, different
instrumentation, different criteria for galaxy selection and different
methods of classifications must be taken into consideration otherwise
results of the comparison may be very misleading.

\subsection{The CFRS-12 sample 
($\bmath{\bld{0.1} < \lowercase{z} < \bld{0.3}}$)} 

The CFRS-12 sample \cite{tre96} is magnitude limi\-ted in the same way
as the SAPM, and has $\langle z \rangle = 0.2$.  Galaxies in CFRS-12
were selected on $I$ magnitude (i.e. more sensitive to old stellar
populations), while the SAPM used $b_J$ selection (i.e. more sensitive
to star forming galaxies).  The galaxy area covered by the 1\farcs75
CFRS slit for galaxies at $z \simeq 0.2$ is similar to the 8\arcsec\
SAPM slit for galaxies at $z \simeq 0.05$ (i.e. $\sim50\%$ of the
galaxy area).

In CFRS-12, $85\%$ of the galaxies are H$\alpha$ emitters with
EW(H$\alpha$ + [\ion{N}{2}]) $\ga 10$ \AA\, and $-21.5 < M(B_{AB}) < -14
$ (see Table~1; Fig.~14 of Tresse et al. 1996).  In
Figure~\ref{plot10}, CFRS-12 galaxies follow the same trend as SAPM
galaxies in Figure~\ref{plot7}; a higher fraction of ELG is found at
fainter blue absolute magnitudes. Similarly to the SAPM, in the
CFRS-12, the non-ELG population is brighter by $\sim1$ mag, and redder
by $\sim0.4$ mag than the ELG population.

To make a fair comparison between the two surveys, we limit the
samples to EW(H$\alpha$ + [\ion{N}{2}]) $\ge 10$ \AA\ for ELG, and
$-17 \leq M(B_{AB} \simeq b_J) \leq -21 $.  We obtain $82\%$ and
$65\%$ of ELG respectively for these subsets of the CFRS-12 and SAPM
samples.  Thus even though SAPM galaxies were selected from their
rest-frame blue continuum, and CFRS-12 galaxies from their rest-frame
red continuum, the fraction of ELG at $z = 0.2$ is higher by a factor
1.3 than locally.  The exclusion of APM merged systems ($< 3\%$, see
Section~2) would not change this result. Moreover the APM galaxy
catalog is known to miss $\sim5\%$ of compact galaxies which are
difficult to distinguish from stars (see Maddox et al. 1990); even if
this fraction is as much as $10\%$, the fractions of ELG for the same
EW limit at low and high redshifts would still be discrepant.

This result is consistent with the rapid evolution in H$\alpha$
luminosity density \cite{tre98a}.  On average at higher redshifts
H$\alpha$ is more intense, and consequently for the same EW limit, the
fraction of ELG increases.  An important point to note is that the ELG
fraction does not depend on the relative normalization of the galaxy
counts, and so this result is an independent demonstration of rapid
galaxy evolution at relatively low redshifts ($z < 0.3$).

\subsection{Other surveys} 

Statistically complete, magnitude-selected spectroscopic surveys which
have H$\alpha$ line information are rare.  In the very nearby Universe
($z \ll 0.005$), Ho et al. \shortcite{ho97b} spectroscopically
observed 486 galaxies from the Revised Shapley--Ames Catalog of bright
galaxies ($B_T \le 12.5$ mag) which contains all morphological types.
Only the central few hundred parsecs of the galaxies have been
observed with a 2\arcsec\ x 4\arcsec\ slit, but the detection limit of
emission lines is very low, at EW $\sim 0.25$ \AA\ ($3\sigma$).  After
removing the stellar background of the bulge, and thus correcting for
stellar absorption (which does not exceed 2--3 \AA, Ho et al. 1997a),
they detected optical line emission in $86\%$ of the galaxy nuclei.
Their result implies that ionized gas is almost invariably present in
the core of galaxies.  From Ho et al. \shortcite{ho97b} figures 4 and
6, we estimate that $\sim$60--65\% of the emission-line nucleus
gala\-xies (ELNG) within roughly the same absolute $B$-magnitude range
as SAPM, have EW(H$\alpha$) $> 2$ \AA.  This fraction is similar to
that found by V\'eron--Cetty \& V\'eron \shortcite{ver86} (60--65\%),
who did not subtract the stellar background.  The differences in
apertures and signal-to-noise mean we cannot directly compare this
fraction to the SAPM or the CFRS-12 samples, however we will make one
comment.

The SAPM 61--62\% ELG fraction is close to the $\sim$60--65\% ELNG.
Since $\sim$40--50\% of the galaxy area is surveyed in the SAPM
galaxies, this suggests that H$\alpha$ emission from the disk causes
little change in the number of H$\alpha$ emitters detected above 2
\AA.  Overall, the nuclear emission amounts to a few percent of the
integrated flux (see Kennicutt \& Kent 1983), so we might have
expected more H$\alpha$ emitters in the case of the SAPM in which the
extra-nuclear emission is observed.  This is not the case, because the
extra-nuclear nebular emission varies roughly in proportion to the
continuum, and thus the EW is relatively unchanged.  This is
consistent with detailed studies by Hunter et al. \shortcite{hun98}
who show that in irregular galaxies, the star formation rate (measured
from H$\alpha$ luminosities) and the total stellar density (measured
from the surface brightness) lie close to each other throughout the
whole galaxy, and by Ryder \& Dopita \shortcite{ryd94} who observed
that the star formation follows the old stellar mass surface density
in spiral galaxies. For the surface brightness profiles of typical
SAPM galaxies, this implies that the detected H$\alpha$ emission is
centrally peaked for most of the SAPM ELG.  This is seen also in other
surveys which select only emission-line galaxies, such as the UCM
survey \cite{gal98}; these find that the line emission comes largely
from the nuclear region.

\section{Emission-line and continuum properties}

We investigate in this section how H$\alpha$, [\ion{N}{2}],
[\ion{S}{2}], [\ion{O}{2}] and [\ion{O}{1}] correlate with $b_J$
luminosity, and D$_{4000}$.

\subsection{H$\alpha$ and M($b_J$)} 

Out of the 990 ELGs, 56 ($6\%$) have detectable H$\alpha$ emission
that could not be properly measured, usually because it was at the
same wavelength as a sky line.  For the remaining 934 ELG, $98\%$ have
EW(H$\alpha$) under 60 \AA, (20 have EW $> 60$ \AA).  The mean and
median EW(H$\alpha$) ($> 2$ \AA) are respectively 19 and 15 \AA.

Figure~\ref{plot142319} shows that the fraction of faint ELG ($M(b_J)
> -21$), increases as a function of EW.  From EW(H$\alpha$) $= 10$ to
60 \AA, the fraction increases by a factor $\sim2$.  Another way to
look at this is to consider the fraction of high EW ELG (EW$ > 15$
\AA), which increases at fainter M($b_J$) (see Fig.~\ref{plot1213}).

These trends are the continuation of those seen for the ELG fraction
as a function of M($b_J$), and they reflect a global tendency within
the whole population; the larger EW(H$\alpha$), the bluer the galaxy
(Kennicutt \& Kent 1983). This is because EW(H$\alpha$) is the ratio
of the flux originating from UV photionization photons ($< 912$ \AA),
over the flux from the old stellar population emitted in the
rest-frame R passband, which forms the continuum at H$\alpha$. Thus
large EW is either due to a large UV flux (or $B$ absolute magnitude
since they are correlated), or to a small continuum from the old
stars. In either case this implies a blue continuum colour.  Hence the
observed trend of larger EW(H$\alpha$) for fainter galaxies implies
that the faint ELG population is dominated by blue galaxies, while the
bright ELG population is dominated by redder galaxies.

So, the galaxies which evolve rapidly with redshift (at least up to $z
= 1$) are faint and dominated by a young stellar population.  The
bright galaxies evolve more or less passively and are dominated by an
older stellar population.  It is now well established that the faint
end of the luminosity function of $B$-selected galaxies is dominated
by actively star-forming galaxies. This is what we find also in the
SAPM \cite{lov99b}.

Since $B$-luminosity is strongly correlated with H$\alpha$ luminosity
\cite{tre98a}, the fainter a galaxy, the smaller its H$\alpha$
luminosity is, and thus the faint population does not dominate the
total H$\alpha$ flux density, or the UV ($< 912$ \AA) ionizing flux
density at low $z$, even though it is the population which evolves
most rapidly with redshift.

\subsection{[N$\;${\sevensize\bf II}], [S$\;${\sevensize\bf II}], 
[O$\;${\sevensize\bf II}], [O$\;${\sevensize\bf I}] detection and
M($b_J$)}

In this subsection, we discuss measurements of [\ion{N}{2}]
$\lambda$6583, [\ion{S}{2}] $\lambda\lambda$6716, 6731, [\ion{O}{2}]
$\lambda$3727 and [\ion{O}{1}] $\lambda$6300 for the 934 H$\alpha$ ELG
with quantified EW(H$\alpha$).  [\ion{N}{2}] $\lambda$6548, which is 3
times fainter than [\ion{N}{2}] $\lambda$6583, has often too low
signal-to-noise to be detected in the same homegeneous way as its 
counterpart at 6583 \AA.   

We measured [\ion{N}{2}] for 784 galaxies ($84\%$).  The average
EW([\ion{N}{2}]) detection is at $4.7\sigma$, which is $\sim19\%$
lower than the one for H$\alpha$.  Since the wavelengths are very
close, the difference is because on average [\ion{N}{2}] has smaller
intensity. This is to be expected since [\ion{N}{2}] is harder to
ionize, and less abundant than H$\alpha$.  The remaining [\ion{N}{2}]
are 46 ($5\%$) with no detection above $\sim2$ \AA, 43 ($5\%$)
detected but too noisy to be measured, 31 ($3\%$) with a sky
subtraction problem, and 30 ($3\%$) not observed because it falls in
the gap at 7000--7020 \AA\ (see Section 2.1).

We measured [\ion{S}{2}] $\lambda\lambda$6716, 6731 for 506 galaxies
($54\%$); for 30 ($3\%$) galaxies only one line could be measured. In
these cases the other line has a too low signal-to-noise to be
reliably measured, and we set the EW to zero. The average
EW([\ion{S}{2}] $\lambda\lambda$6716, 6731) detection for these 536
galaxies is at $4.3\sigma$, which is $\sim30\%$ lower than
H$\alpha$. This is partly because [\ion{S}{2}] has a smaller intensity
than H$\alpha$ (as in the case of [\ion{N}{2}]), but also because the
continuum at 6717--6731 \AA\ is more noisy than at H$\alpha$; sky
lines start to increase the noise in the red continuum.  For 123
galaxies no line was detected, and for the remaining 275 galaxies, one
of the lines was either on a sky line or in the gap. The percentage of
[\ion{S}{2}] detected above 2 \AA\ is higher than for [\ion{N}{2}],
because the sum has higher intensity, even though individual lines are
harder to detect.

We measured [\ion{O}{1}] $\lambda$6300 for 43 galaxies (3\%).  The
average detection of EW[\ion{O}{1}] is 2.7$\sigma$, it is the lowest
value since this line intensity is usually extremely weak relatively
to the other lines.

We measured [\ion{O}{2}] for 859 galaxies ($92\%$).  The average
detection is $4.8\sigma$, which is $\sim15\%$ lower than for
EW(H$\alpha$). This is because the blue continuum is less intense than
the red continuum, and consequently more noisy.  We find 75 galaxies
with significant H$\alpha$ but no detectable [\ion{O}{2}].  The
EW(H$\alpha$) for these galaxies is detected at more than $3\sigma$,
and lies in the range 2--24 \AA, with a mean at 8 \AA.  They have also
$1.3 < D_{4000} < 1.8 $, and a median $D_{4000} = 1.5$. Out of the 75,
58 have [\ion{N}{2}] detected, and 15 have also [\ion{S}{2}] detected.
These galaxies are undergoing star bursts, and are probably heavily
absorbed so that they show a moderate Balmer break, and a weak noisy blue
continuum (see also Section 8). They are not preferentially edge-on
galaxies.

Figure~\ref{plot142319} shows that the fraction of faint galaxies
($M(b_J) > -21$) increases as a function of: (a) [\ion{N}{2}] EW by a
factor 1.2 from 2 to 15 \AA, (b) [\ion{S}{2}] EW by a factor 2 from 2
to 20 \AA, and (c) [\ion{O}{2}] EW by a factor 2 from 2 to 40 \AA.
These trends are analogous to what we see for H$\alpha$, where the
fraction of faint ELG increases by a factor 2 up to EW(H$\alpha$) $=
60$ \AA.  Faint galaxies have larger EW for emission lines both in the
blue and red band at rest.  Thus, as well as being bluer as seen
previously, they have also stronger photoionization sources which
enhance forbidden line intensities.  Indeed, if there was simply a
larger amount of young stellar background in the blue, then
EW([\ion{O}{2}]) would decrease while EW([\ion{S}{2}]) would increase,
(noting that [\ion{O}{2}] has similar photoionization conditions to
[\ion{S}{2}]).  So, the enhancement of UV photoionization sources is
coupled with a larger amount of $B$-band flux produced mainly by
intermediate-mass stars. This leads to the correlation between
H$\alpha$ and $B$ luminosity \cite{tre98a}.  These metallic lines
follow the same trend as H$\alpha$, suggesting that they have a common
source of ionization.  The small number of galaxies with detected
[\ion{O}{1}] makes the comparison more difficult, however the fraction
of faint galaxies seems not to be correlated with this weak line.
Emission-line ratios are studied in the following Section, and
correlations are studied in Section~7.

The medians of the EW distributions for the bright and faint ELG
populations are listed in Table~\ref{table4}.  The difference between
these medians is largest for [\ion{O}{2}] (47\%), then H$\alpha$ and
[\ion{S}{2}] ($\sim 31$\%).  We can interpret these variations as
follows: EW([\ion{O}{2}]) is the most sensitive to blue luminosity,
since it depends directly on the rest-frame $B$-band continuum;
EW(H$\alpha$) and EW([\ion{S}{2}]) are slightly less sensitive since
they depend on the rest-frame $R$-band continuum.

For [\ion{N}{2}] there is no significant difference (4\% discrepancy).
The global behavior of EW([\ion{N}{2}]) reflects intrinsic differences
in the nitrogen abundance (N/O) in ELG. The origin of nitrogen is
still an open debate (see for instance data analysis in Izotov \&
Thuan \shortcite{iso99}, Garnett \shortcite{gar90}, Kobulnicky \&
Skillman \shortcite{kob98} and references within). On one hand, the
nature of stars producing the primary nitrogen (i.e. from burning H
and He via fresh C and O) remains unclear in low-metallicity systems
(high- or intermediate-mass stars). On the other hand, the scenario to
explain a second source for nitrogen enrichment observed in
high-metallicity galaxies is still open.  We cannot address these
points in detail here. However in our representative local sample, we
find that on average luminous blue ELG are likely to be enhanced in
nitrogen abundance.  This suggests that in faint, low-mass, late-type
ELG, nitrogen is a primary element, whereas in more bright, more
massive galaxies nitrogen it comes from a secondary source. Thus the
expected global increase in the fraction of faint galaxies as a
function of line strengths would not be observed in the case nitrogen
enhanced in a non negligible fraction of bright galaxies.

\subsection{Emission-line EW and D$_{4000}$}

As seen in Section~3 the presence of H$\alpha$ in emission is
correlated with a low D$_{4000}$ index.  This trend continues within
the ELG population; the larger is EW(H$\alpha$), the smaller is
D$_{4000}$.

Figure~\ref{break6} shows that the fraction of ELG with low D$_{4000}$
($< 1.4$) increases as a function of: (a) H$\alpha$ EW by a factor 4.5
from 2 to 60 \AA, (b) [\ion{O}{2}] EW by a factor 3.4 from 2 to 40
\AA, (c) [\ion{S}{2}] EW by a factor 4.6 from 2 to 20 \AA, (d)
[\ion{N}{2}] EW by a factor 2 from 2 to 15 \AA.  The medians of the EW
distributions for the low and high D$_{4000}$ ELG populations are
listed in Table~\ref{table4}.  They are the most discrepant for
H$\alpha$ (69\%), then [\ion{O}{2}] (52\%), [\ion{O}{1}] (40\%),
[\ion{S}{2}] (36\%), and [\ion{N}{2}] (35\%).  EW(H$\alpha$) is the
most sensitive to D$_{4000}$ as expected, since these two measurements
are the most sensitive tracers of star formation rate.  Here
[\ion{N}{2}] and [\ion{O}{1}] follow the same trend as for the other
lines, in contrast to their behavior as a function of rest-frame blue
luminosity. This is because a low D$_{4000}$ is caused by the
high-mass stars which enhance all line strengths.

\section{Emission-line ratios}

\subsection{[N$\;${\sevensize\bf II}]/H$\alpha$, 
[S$\;${\sevensize\bf II}]/H$\alpha$, 
[O$\;${\sevensize\bf I}]/H$\alpha$ ratios}

The ratio [\ion{N}{2}] $\lambda$6583/H$\alpha$ involves lines close in
wavelength, and thus it does not depend on reddening or flux
calibration.  The EW ratio must be in principle equivalent to the flux
ratio, since the continuum is the same. It is a good ratio to compare
galaxies at different redshifts.

Only substantial stellar absorption at H$\alpha$ may affect this
comparison: in late-type galaxies, it is usually negligible; it may
become significant in early types, which exhibit Balmer absorption
lines.  For SAPM galaxies where [\ion{N}{2}] is detected, it is
certainly negligible.  Indeed only detai\-led studies with a level of
detection lower than 2 \AA\ EW can detect weak [\ion{N}{2}] in
early-type galaxies \cite{ho97a}.

Figure~\ref{plot15b} shows the distribution of the
[\ion{N}{2}]/H$\alpha$ EW ratios for our 784 galaxies with
EW(H$\alpha$) + 1.33EW([\ion{N}{2}] $\lambda$6583) $> 3$ \AA.  We note
that the ratio does not show any trend with redshift.  The median and
mean values are 0.37 and 0.41, and we find much the same values using
our relative flux ratios, 0.36 and 0.40.  We also show the
distribution of the ratios published by Kennicutt (1992, hereafter
K92) in his tables 1 \& 2 respectively of high (5--7 \AA) and low ($>
10$ \AA) resolution spectra.  We excluded the \hbox{H\,{\sc ii}}
regions (Mkr~59 and Mkr~71), the Seyfert~1 galaxies (NCG~3516,
NGC~5548 and NGC~7469) and galaxies with EW(H$\alpha$) +
EW([\ion{N}{2}] $\lambda\lambda$6748, 6583) $< 3$ \AA\ as in the SAPM.
We note that his table~2 contains galaxies already tabulated in
table~1. For these, we used the measurements from his table~1.  In
total, K92's sample has 57 ratios of narrow-line fluxes integrated
over the whole galaxies.  Since the K92 galaxies are not a complete
magnitude-limited sample, it is dan\-gerous to consider the observed
distributions of any parameter as representative of the true
distributions.  Therefore we calculated the distribution of the
[\ion{N}{2}]/H$\alpha$ EW ratios separately for each morphological
type, and averaged them with weights proportional to the fraction of
each type in the RSA (Sandage \& Tammann 1981). The median and mean
ratios are 0.38 and 0.47 for this K92 sample.

The median and mean ratios for the SAPM and K92 samples are
similar. Kennicutt's values are slightly higher because his ratios are
likely to be biased towards higher values, as discussed in K92.  This
tells us that the SAPM spectral properties are on average well
representative of integrated spectra (see also Fig.~\ref{plot30}).  As
noted by K92, undersampling the disk tends to reduce the strengths of
the emission lines in roughly equal proportion, and thus the rela\-tive
line fluxes should be insensitive to it. This strengthens our
conclusions from Section~2, showing that our results are very unlikely
to be affected or biased by using a long slit. 

The median and mean for the EW ratio, [\ion{S}{2}] $\lambda\lambda$
6716, 6731/H$\alpha$, are 0.36 and 0.42, and for the flux ratio, are
0.36 and 0.41.  If we take only our spectra for which [\ion{S}{2}]
lines fall on the same CCD chip as H$\alpha$, we have similar values,
respectively 0.37 and 0.43, 0.36 and 0.41. This tells us that the
relative calibration for the continuum of the two red chips is on
average good to about $1\%$.  The median and mean for the EW ratio,
[\ion{O}{1}] $\lambda$ 6300/H$\alpha$, are 0.08 and 0.11, and for the
flux ratio, are 0.08 and 0.42. The fact that the EW and flux ratio
averages are similar shows that the red continuum at H$\alpha$ and at
[\ion{S}{2}] or [\ion{O}{1}] is not significantly different. 

[\ion{N}{2}]/H$\alpha$, [\ion{S}{2}]/H$\alpha$ and
[\ion{O}{1}]/H$\alpha$ (Fig.~\ref{plot15b}) are commonly used to
distinguish galaxies hosting an AGN from the others. We analyse them
elsewhere.

\subsection{[N\,{\sevensize\bf II}]/H$\alpha$ versus 
EW(H$\alpha$ + [N\,{\sevensize\bf II}])}
 
In deep optical surveys, H$\alpha$ and [\ion{N}{2}]
$\lambda\lambda$6548, 6583 lines are often blended because of the use
of low-resolution spectroscopy. It is however important to recover the
flux solely in H$\alpha$ to measure for instance the H$\alpha$
luminosity function, hence to derive a star formation rate. This is
also necessary to distinguish AGN galaxies from \hbox{\ion{H}{2}}
galaxies, in particular in narrow-line emission galaxies.  Broad-line
galaxies are identified straightforwardly as AGN, and they are not
numerous in representative surveys of field galaxies.

The value of the ratio [\ion{N}{2}] $\lambda\lambda$6548,
6583/H$\alpha$ is usually taken to be 0.5 to remove the contribution
of [\ion{N}{2}] to (H$\alpha$ + [\ion{N}{2}]) blended
emission, as determined by Kennicutt \shortcite{ken92}.  Using the
SAPM sample, we study this ratio in more detail.  We note that
including AGN changes the average values very little, since the 
overall emission is dominated by stellar emission in local
representative surveys.

Figure~\ref{plot18new}a shows that [\ion{N}{2}]
$\lambda$6583/H$\alpha$ EW ratio decreases as a function of
EW(H$\alpha$).  In this plot we consider only spectra with [N\,{\sc
ii}] EW detected above $3\sigma$.  We fitted a least square relation
to the log of the data ($\log $[\ion{N}{2}]/H$\alpha = -0.4 \log
$EW(H$\alpha$) $+ 0.1$).  At our median EW(H$\alpha$), the relation
gives a [\ion{N}{2}]/H$\alpha$ ratio of 0.4, i.e. 0.5 for
([\ion{N}{2}] $\lambda\lambda$6548, 6583)/H$\alpha$.  We recall that
1.33~[\ion{N}{2}] $\lambda$6583 should be equi\-valent to [\ion{N}{2}]
$\lambda\lambda$6548, 6583 since [\ion{N}{2}]
$\lambda$6548/[\ion{N}{2}] $\lambda$6583 $= 1/3$.  In this figure, we
also plot K92's sample; we see that the high-resolution K92 subsample
is generally above our SAPM trend because it contains a large fraction
of early-type galaxies which have systematically higher ratios.  The
low resolution K92 data are slightly higher than our median value
because of K92's exclusion of weak H$\alpha$ + [\ion{N}{2}] blended
lines.

Figure~\ref{plot18new}b shows the relation 1.33
[\ion{N}{2}]/H$\alpha$ EW versus EW(H$\alpha$) +
1.33~EW([\ion{N}{2}]).  The trend is the same as in
Fig.~\ref{plot18new}a.  Thus we can predict which value is expected
for the ratio when observing the blend H$\alpha$ + [\ion{N}{2}].  For
instance, if this latter is $\sim100$ \AA, the ratio should be
$\sim0.3$, whilst if it is $\sim20$ \AA, it should be $\sim0.7$.

In terms of galaxy numbers, the slight overestimation for bright
galaxies (mainly low EW), and underestimation for faint galaxies
(mainly high EW) are about counter\-balanced, since 0.5 is the median
value.  In fact, as we will see in Section~7, our relations
[\ion{O}{2}]--H$\alpha$ and [\ion{O}{2}]--(H$\alpha$ + [\ion{N}{2}])
are equivalent if this median is taken.  However this argument applies
only for a similar EW(H$\alpha$) distribution.  Samples at higher
redshifts are skewed towards higher EW compared to the local samples
(at least in [\ion{O}{2}]~EW, see Fig.~\ref{plot30}), and thus the
local median, $0.5$, is not the correct one to use. This effect will
be more significant in small samples at high redshift.

\section{Correlation between emission lines}

\subsection{[N\,{\sevensize\bf II}], [S\,{\sevensize\bf II}], 
[O\,{\sevensize II}],  [O\,{\sevensize I}]  versus H$\alpha$}

Figs.~\ref{plot25}, ~\ref{plot25a}, ~\ref{plot25b} \& ~\ref{plot25e}
show that EW of [\ion{N}{2}] $\lambda$6583, [\ion{S}{2}]
$\lambda\lambda$6716, 6731, [\ion{O}{2}] $\lambda$3727, [\ion{O}{1}]
$\lambda$6300 increase as a function of H$\alpha$ EW.  The figures are
on a log-log scale to show all points, in particular those with low
EWs. We note that this scale enhances the dispersion of low EWs in
comparison with the one of high EWs.  Table~\ref{table3} gives the
correlation parameters.  The correlations are measured with EW, and
not log(EW).

The large scatter (rms of about $50\%$) found in the EW correlations
is not due to poor signal-to-noise; the scatter is still large even
with lines detected above $5\sigma$.  In the case of [\ion{N}{2}] or
[\ion{S}{2}] lines, it is independent of the red stellar background.
It simply reflects real scatter in the main physical parameters
(metallicity, effective temperature, ionisation parameter) which drive
variations in the strengths of various forbidden transitions relative
to the recombination lines. In addition, any presence of AGN, in
particular in early Hubble types, and any contribution to the emission
from the diffused ionized gas will increase the primary scatter (see
for instance \cite{ho97c} and references within).  The dispersion is
larger for the [\ion{O}{2}] correlation.  In this case, the effect is
accentuated by the diversity of young stellar contents which produce
the continuum at [\ion{O}{2}]. Although the [\ion{O}{1}] line is
rarely detectable, we were able to measure the EW in a few
galaxies. For this line the dispersion is as large as $\sim$80\%. The
scatter is so large partly because of the weakness of the line
intensity which leads to low signal-to-noise, and partly because most
of the [\ion{O}{1}] flux comes from the partially ionized transition
zone produced by high-energy photoionisation, which means the
[\ion{O}{1}]/H$\alpha$ ratio is very sensitive to the structure and
thickness of the zone.  We flagged all objects having [\ion{N}{2}]
$\lambda$6563/H$\alpha > 0.63$ (i.e. as good candidates for hosting an
AGN) with starred symbols. Excluding them changes the correlations by
less than $5\%$.  The fractional dispersions are almost independent of
the EW strength. Since EW is correlated with luminosity (see Section
5.1), this implies that both faint and bright galaxies have a variety
of photoionization environments.

Since [\ion{O}{2}] is at short optical wavelengths and is correlated
with H$\alpha$ strength, it has been used as a star formation rate
(SFR) indicator for high-$z$ galaxies, when H$\alpha$ is not visible
in the optical window at $z >$ 0.3--0.4.  From the K92 relation,
EW([\ion{O}{2}]) $= 0.4$ EW(H$\alpha$ + [\ion{N}{2}]
$\lambda\lambda$6548, 6583), we assumed his quoted value [\ion{N}{2}]
$\lambda$6583/H$\alpha$ = 0.53, and derived EW([\ion{O}{2}]) $\simeq
0.6$ EW(H$\alpha$).  Our value is discrepant with K92 by $\sim10\%$,
which is not significant given the dispersion of the data in both
samples.

We note that EWs measured from the overall galaxy content are likely
to be affected by dust. Indeed, recombination emission lines originate
from dusty \hbox{\ion{H}{2}} regions where hot stars (OB type) are
formed, whereas the continuum at [\ion{O}{2}] or H$\alpha$ comes from
long-lived stars sitting in less, or non obscured
regions. EW([\ion{O}{2}]) is more affected by the dust than
EW(H$\alpha$). Hence, the intrinsic [\ion{O}{2}]--H$\alpha$ EW
ratio should be on average slightly larger than our observed
relation.

\subsection{The [O\,{\sevensize\bf II}]--(H$\alpha$ + [N\,{\sevensize\bf II}]) 
relation in different surveys}

The correlation between EW([\ion{O}{2}]) and EW(H$\alpha$ +
[\ion{N}{2}] $\lambda\lambda$6548, 6583) has been commonly used to
estimate SFR.  In Figure~\ref{plot25c}, we show it for our SAPM data,
and in Figure~\ref{plot25d} we plot it together with data from the K92
and CFRS-12 samples.  The distribution of K92 data is spread over the
SAPM distribution; the correlation and dispersion of K92 data are
similar to the SAPM with as little as 5--10\% discrepancy (see
Table~\ref{table3}).  This is just about within the expected random
noise; the rms in the mean relation is $\sim8\%$ for K92, and
$\sim$2--3\% for the SAPM.

K92 sample has a larger fraction of high EW data.  High EWs in K92
sample usually correspond to late Hubble type galaxies (see fig. 11 in
Kennicutt 1992).  Our median for [\ion{O}{2}] and H$\alpha$ +
[\ion{N}{2}] are about half those for the K92 data.  Our
EW([\ion{O}{2}]) distribution at $z = 0.05$ is very similar to nearby
samples (see Fig.~\ref{plot30}), except for EW $< 5$ \AA, which is
pro\-bably due to different detection levels, and magnitude ranges
(luminous galaxies of the local Universe usually have low EW).  Thus
this factor two cannot be due to our undersampling, but rather
reflects the difference between a representative sample (SAPM) and a
sample selecting specific galaxies (K92).

CFRS-12 data, which is a small representative sample at $\langle z
\rangle = 0.2$ of galaxies usually fainter than $-21$ in $B$, have
stronger EWs. On one hand, the small number of low EW is certainly due
to the poor detection of EW lower than 10~\AA\ in the CFRS spectra,
and the lack of luminous galaxies at $z < 0.3$. On the other hand, the
larger fraction of strong EW spectra is genuine, and corresponds to
the rapid evolution of faint galaxies, as discussed previously.  The
distribution of EW in distant surveys such as the CFRS is
significantly different than to the local ones (see
Fig.~\ref{plot30}).  Another observation is that the
[\ion{O}{2}]/(H$\alpha$ + [\ion{N}{2}]) ratio for the CFRS-12 galaxies
(see Table~\ref{table3}) is $\sim40\%$ higher than in the local
Universe, but the scatter in the data is also larger. The ratio is
higher because spectra exhibit stronger [\ion{O}{2}] EW than local
galaxies with same H$\alpha$ EW. This has been noted by Hammer et
al. \shortcite{ham97}.  Since the CFRS-12 flux ratio is exactly the
same as the CFRS-12 EW ratio, then differences in the color of the
continuum do not change the relation. Even strong stellar absorption
at H$\alpha$ would not be sufficient to explain the shift towards
smaller H$\alpha$ EWs than expected from the local relation at strong
[\ion{O}{2}] EWs.  Therefore the apparent excess of abnormally strong
[\ion{O}{2}] must correspond to a genuine change in faint galaxies
towards higher redshifts, which produces larger degrees of
ionization. This suggests that the local relation between [\ion{O}{2}]
and H$\alpha$ should not be extrapolated to distant galaxies with
strong [\ion{O}{2}] and very blue continuum, which are more numerous
than locally. For these galaxies, the SFR estimated from the
[\ion{O}{2}] EW and the local relation will be overestimated by as
much as $\sim50\%$. Clearly amongst emission lines, H$\alpha$ is the
most reliable tracer of SFR for distant galaxies (see Tresse, Maddox
\& Loveday 1998).

\section{[O\,{\sevensize\bf II}] detected in non-ELG}

Out of 599 galaxies with no H$\alpha$ detected, 68 ($11\%$) exhibit
[\ion{O}{2}] $\lambda$3727.  Their distribution is represented in
Figure~\ref{plot30}; their EW are mainly lower than 10 \AA. They are
not found at a particular redshift, or magnitude. They have $D_{4000}
> 1.5$, with a median at 1.8, i.e. these [\ion{O}{2}] emitters are
lying prefe\-rentially in red galaxies.  According our relation
between H$\alpha$ and [\ion{O}{2}], they should have EW(H$\alpha$)
$\simeq 15$ \AA. Only strong stellar absorption ($> 10$ \AA) in these
red objects would not allow us to detect H$\alpha$.  We note that the
distribution of their EW([\ion{O}{2}]) does not have a preferred
galaxy inclination.  They are most likely early Hubble type galaxies
undergoing modest starbursts. Their morphology is as follows: 10
elliptical, 10 SO, 39 spirals, 1 irregular, and 5 unclassified.  They
represent only $4\%$ of the total SAPM sample. Analogous objects have
been found in the CFRS at $z > 0.45$ (see Hammer et al. 1997),
i.e. [\ion{O}{2}] detected with red colors (H$\alpha$ is not visible).
A fraction of $6\%$ (13 out of 210) are detected with an old stellar
background. These CFRS galaxies seem to be the equivalent of our
non-ELG with [\ion{O}{2}] lines detected. It seems that the fraction
of such objects increases with redshift, indeed the CFRS has a much
lower spectral resolution than the SAPM, which increases the detection
limit of [\ion{O}{2}], and certainly more [\ion{O}{2}] emitters within
these early-type galaxies should be found. 

Another $5\%$ (10 out of 210) in the CFRS are detected with a young
stellar background, but heavily reddened. These galaxies seem more
likely to correspond to our galaxies with H$\alpha$ but no
[\ion{O}{2}] detected (see Section 5.2). In our SAPM spectra,
[\ion{O}{2}] is not always detected with certainty because the blue
continuum is too noisy for these galaxies. In addition, the median of
the [\ion{O}{2}] EW distribution of local galaxies is lower by a
factor $>$2 than the one for high-$z$ distributions, and so EW are
expected to be on average smaller (see Table~\ref{table5}).  They
represent $4\%$ (75 out of 1671) of the total SAPM sample, and thus
their fraction seems to be constant with redshift.

If we classify the SAPM galaxies according their [\ion{O}{2}]
emission, then we find $\sim 61$\% of [\ion{O}{2}] emitters, which is
the same fraction as H$\alpha$ emitters. Indeed galaxies with
[\ion{O}{2}] but no H$\alpha$ are counterbalanced by galaxies with
H$\alpha$ and no [\ion{O}{2}].

\section{Conclusion}

In this paper we have analysed various line EWs and the 4000 \AA\
Balmer break of galaxies in the SAPM survey. This survey uniformly
samples all galaxy types, and the spectral properties are
representative of the local Universe at $z \sim 0.05$.  We would like to
point out in particular that our results are independent of any
incompleteness in the SAPM morphological classifications and
uncertainties in the SAPM flux calibration.  Thus our results can on
average be compared with deeper surveys.  Our main results are: 
\begin{enumerate}
\item $61\pm1\%$ of the SAPM galaxies are H$\alpha$-emitters (ELG) with
EW(H$\alpha$) $\ga 2$ \AA.  The detection of H$\alpha$ in emission
indicates the presence of newly-formed, short-lived stars ($t<$ few
$10^6$ yr) which radiate at $\lambda < 912$ \AA. The ELG fraction
increases as the galaxy is fainter and physically smaller.  This
fraction is larger in high and low surface brightness galaxies,
i.e. in compact and dwarf galaxies.  ELG have low 4000 \AA\ Balmer
breaks, i.e. consistent with massive star formation.  Faint ($L <
L^*_B$) galaxies are bluer, and have stronger photoionization sources
than bright galaxies.  This agrees with the general picture where
faint and small galaxies are more actively forming stars than their
bright counterparts in the local Universe.  This is reflected in the
difference between ELG and non-ELG luminosity functions
\cite{lov99b}. It is also consistent with a ``peak'' in the SFR history
since the overall luminosity density is dominated by bright galaxies,
which formed most of their stars much earlier than faint galaxies,
which are actively forming stars today.  The population of small, less
massive systems is what evolves rapidly, and undergoes rapid
brightening towards earlier epochs. \\
\item Comparison of the ELG fraction with the 
CFRS-12 sample ($\langle z \rangle = 0.2$) demonstrates
a rapid evolution of the faint galaxy population at low redshifts.  This
observation does not depend on the relative normalization of the
galaxy counts. Emission lines trace the massive, young, short-lived
star galaxy contents, and thus their evolution is more rapid than
color evolution.  However, since the SFR follows the total stellar or
gas density, no strong change in the average spectral properties
should be detected, unless some process rapidly enhances the SFR.
Except in the case of interacting galaxies, where disruption of the
galaxy density will provide a good site for starbursting, for the
remaining population, other factors must be taken into account. \\
\item We note a continuity in the spectral properties and luminosities
of ELG and non-ELG, and of strong and weak H$\alpha$ emitters.  In
addition, a small D$_{4000}$ is closely related to the presence of H$\alpha$
in emission, and to the EW strengths of recombination and forbidden
lines.  We do not identify a special class of galaxies, except those
$4\%$ galaxies discussed in (vii) below. \\
\item The ratio [\ion{N}{2}] $\lambda$6583/H$\alpha$ decreases with
EW(H$\alpha$).  Its median is 0.5 for nearby galaxies as found by
Kennicutt \shortcite{ken92}. However this value should only be applied
to the H$\alpha$ + [\ion{N}{2}] $\lambda\lambda$6548, 6583 blend for
galaxies with similar EW(H$\alpha$) distributions to local ones and is
not appropriate for high-$z$ samples. \\
\item{ [\ion{O}{2}], [\ion{S}{2}], [\ion{N}{2}], [\ion{O}{1}] and
H$\alpha$ are all correlated, but present large dispersions
($\sim50\%$, and even larger for [\ion{O}{1}]), which reflect the
diversity in the photoionisation processes. The relation
between [\ion{O}{2}] and H$\alpha$ EW of the SAPM galaxies gives on
average a SFR smaller by $\sim10\%$ than the K92 derived
relation. Moreover this relation seems to change at high redshifts,
and the distribution of [\ion{O}{2}] EWs evolves with redshift.  Thus
SFRs estimated from [\ion{O}{2}] EWs and the local relation may be
overestimates.  In particular, using [\ion{O}{2}] for the faint, blue
galaxies, the individual SFR may be overestimated by as much as
$\sim50\%$. H$\alpha$ remains the most reliable indicator for distant
galaxies.} \\
\item On average luminous blue ELG are likely to be enhanced
in nitrogen abundance.  This suggests that in faint, low-mass,
late-type ELG, nitrogen is a primary element, whereas in more bright,
more massive galaxies nitrogen it comes from a secondary source. \\
\item Only $4\%$ of non-ELG (H$\alpha$ not detected) have [\ion{O}{2}]
detected, and correspond to early-type galaxies undergoing modest
starburts.  Their fraction seems to increase with redshift. 
\end{enumerate}

\section*{Acknowledgments}
We thank the referee L. Ho for his careful reading of the paper.

\bsp 

\clearpage

\begin{table*}
\caption{Fractions $F(a)$ and $F(b)$ covered by a 8\arcsec\ slit for
different ranges of ellipticities, $S_{25}$ and $z$, as described in
Section 2.3. \label{table1}}
\begin{tabular}{cccc} 
Ellipticity & N    & F(a) &  F(b) \\ 
  range  $\  \  \ \ $ mean  &  & mean & mean \\ 
\tableline
\hfill \\
0.00--1.00$\  \  $ 0.62  & 1671 (ALL) &   57\% &  33\%  \\ 
0.00--0.64$\  \  $ 0.44  &  836       &   67\% &  28\% \\ 
0.64--1.00$\  \  $ 0.80  &  835       &   47\% &  37\% \\
\hfill \\
Log S$_{25}$ (kpc$^2$) & N    &  F(a) & F(b)  \\ 
  range  $\  \  \ \ $ mean  &  & mean & mean \\ 
\tableline
\hfill \\ 
0.80--4.00$\  \  \ \ $  2.88 & 1671 (ALL)  &  57\% & 33\%  \\
0.80--2.94$\  \  \ \ $  2.56 & 836         &  62\% & 33\%  \\
2.94--4.00$\  \  \ \ $  3.19 & 835         &  51\% & 32\%  \\
\hfill \\
z & N    &  F(a) & F(b)  \\ 
  range  $\  \  \ \ $ mean  &  & mean & mean \\ 
\tableline
\hfill \\ 
0.0036--0.1422$\  \  \ \ $  0.0531 & 1671 (ALL)  &  57\% & 33\%  \\
0.0036--0.0533$\  \  \ \ $  0.0346 & 836         &  56\% & 29\%  \\
0.0533--0.1422$\  \  \ \ $  0.0717 & 835         &  58\% & 36\%  \\
\hfill \\
\end{tabular}
\end{table*}

\begin{table*}
\begin{minipage}{150mm}
\caption{Average parameters and respective $rms$ (see Section~3.1 for details of the classification)  
\label{table2}}
\begin{tabular}{lllllllll}
Average parameters &  ELG & ($rms$)  &  non-ELG & ($rms$) & unclassified & ($rms$) & ALL & ($rms$)    \\
\tableline
\hfill \\
Number           &  990 &  & 599 & & 82&  & 1671&  \\
\hfill \\
$b_J$\ [mag]    &  16.53 & (0.02)  & 16.46&(0.02) &  16.58 &(0.05)   & 16.51&(0.01) \\
$z$               & 0.0504& (0.0008) & 0.0559&(0.0009) & 0.0663&(0.0003) & 0.0531&(0.0006) \\
M($b_J$)\ [mag] & $-$20.73 & (0.04) & $-$21.16&(0.04) & $-$21.65&(0.06) & $-$20.93&(0.03) \\ 
$k$\ [mag]       &  0.142 & (0.002) & 0.173&(0.004) & 0.200&(0.009)  & 0.156&(0.002) \\
Log $S_{25}$\ [kpc$^{2}$] & 2.79 & (0.01) & 2.99& (0.02) & 3.14&(0.02) & 2.88&(0.01) \\
$\mu_{25}$\ [mag arcsec$^{-2}$] & 22.81& (0.01) & 22.89&(0.01) & 22.76&(0.03) & 22.839& (0.007) \\ 
D$_{4000}$      & 1.395&(0.006) & 1.83 & (0.01) & 1.65 & (0.03) & 1.563& (0.008) \\
 $i$ [deg] & 52.3 &(0.6)  & 51.1 &(0.7) & 48.7 & (0.7) & 51.7& (0.4) \\
\hfill \\ 
\tableline
\end{tabular}

\medskip
$b_J$ is the apparent magnitude; $z$ is the redshift; M($b_J$) is the
absolute magnitude; $k$ is the $k$-correction for $b_J$ magnitudes,
calculated individually for different morphological types as used by
Loveday et al. (1992); $S_{25}$ is the rest-frame projected area
brighter than a $b_J$ surface brightness level of 25 mag
arcsec$^{-2}$; $\mu_{25}$ is the rest-frame surface brightness
averaged over the area brighter than $b_J = 25$ mag arcsec$^{-2}$;
D$_{4000}$ is the Balmer index at 4000 \AA; and $i$ is the galaxy
inclination.
\end{minipage}
\end{table*}

\begin{table*}
\begin{minipage}{150mm}
\caption{Median and number of galaxies of the EW distributions for
bright and faint ELG, and for low and high $D_{4000}$ ELG (see also
Fig.~\ref{plot142319}, and Fig.~\ref{break6}).
\label{table4}}
\begin{tabular}{lrrrrr}
 & EW(H$\alpha$) & EW([\ion{O}{2}]3727) & EW([\ion{S}{2}]6717, 6731) & EW([\ion{N}{2}]6583) & EW([\ion{O}{1}]6300) \\
\tableline
\hfill \\
ALL                       & 14.9 [934] & 10.4 [859] & 6.9 [536] & 5.7 [784] & 2.0 [43] \\
\hfill \\
(L $\ge$ L$^{*}_B$) ELG   & 12.7 [417] & 8.1  [371] & 5.4 [193] & 5.6 [350] & 1.9 [17] \\
(L $<$ L$^{*}_B$) ELG     & 17.5 [517] & 13.0 [488] & 7.5 [343] & 5.8 [434] & 2.0 [26] \\ 
\hfill \\
(D$_{4000}$ $\ge$ 1.4) ELG   & 10.6 [467] &  8.1 [409] &  5.3 [234] & 4.7 [394] & 1.7 [20]  \\
(D$_{4000}$ $<$ 1.4 ) ELG    & 20.9 [467] &  13.5 [449] & 7.8 [302] & 6.7 [390] & 2.5 [23]  \\ 
\hfill \\
\tableline
\end{tabular}
\end{minipage}
\end{table*}

\begin{table*}
\begin{minipage}{150mm}
\caption{Correlations between emission-line EW \label{table3}}
\begin{tabular}{rclccccc} 
\tableline 
\multicolumn{3}{c}{Median correlation} &  & rms & & \multicolumn{2}{c}{Median}\\
EW(2) & = & a EW(1) &  EW $\sigma$ level$^1$ & dispersion & N & EW(1) & EW(2)   \\
\tableline
\hfill \\ 
\, [\ion{N}{2}] $\lambda$6583 & = & 0.37 H$\alpha$ & any & $\sim55$\% & 784  & 15.6 &  5.7  \\
\, [\ion{N}{2}] $\lambda$6583 & = & 0.41 H$\alpha$ & 3$\sigma$& $\sim50$\%  & 553   & 15.8  &  6.5  \\
\, [\ion{N}{2}] $\lambda$6583 & = & 0.45 H$\alpha$ & 5$\sigma$& $\sim42$\%  & 280   & 17.8  &  7.8  \\
\hfill \\
\, [\ion{S}{2}] $\lambda\lambda$6716, 6731 & = & 0.36 H$\alpha$ & any & $\sim57$\% & 536  & 18.1 & 6.9 \\
\, [\ion{S}{2}] $\lambda\lambda$6716, 6731 & = & 0.39 H$\alpha$ & 3$\sigma$ & $\sim48$\% & 361  & 20.0 & 7.8 \\
\, [\ion{S}{2}] $\lambda\lambda$6716, 6731 & = & 0.40 H$\alpha$ & 5$\sigma$ & $\sim46$\% & 143  & 24.4 & 10.0 \\
\hfill \\
\, [\ion{O}{2}] $\lambda$3727 & = & 0.67 H$\alpha$  & any & $\sim68\%$ &  859  & 15.6 & 10.4 \\
\, [\ion{O}{2}] $\lambda$3727 & = & 0.69 H$\alpha$  & 3$\sigma$  & $\sim64\%$ &  656 & 17.0 & 11.5 \\
\, [\ion{O}{2}] $\lambda$3727 & = & 0.67 H$\alpha$  & 5$\sigma$  & $\sim57\%$ &  294 & 21.6 & 14.6 \\
\hfill \\
\, [\ion{O}{2}] $\lambda$3727 & = & 0.45 (H$\alpha$ + [\ion{N}{2}] $\lambda\lambda$6548, 6583)  & any & $\sim65\%$ &  769  & 24.2 & 10.6 \\
\, [\ion{O}{2}] $\lambda$3727 & = & 0.47 (H$\alpha$ + [\ion{N}{2}] $\lambda\lambda$6548, 6583)  & 3$\sigma$  & $\sim59\%$ &  593 & 26.3 & 11.7 \\
\, [\ion{O}{2}] $\lambda$3727 & =  & 0.47 (H$\alpha$ + [\ion{N}{2}] $\lambda\lambda$6548, 6583)  & 5$\sigma$  & $\sim52\%$ &  280 & 32.5 & 14.7 \\
\hfill \\
\, [\ion{O}{2}] $\lambda$3727 & =  & 0.42 (H$\alpha$ + [\ion{N}{2}] $\lambda\lambda$6548, 6583)  & K92 data$^2$  & $\sim61\%$ & 63  &  60 & 25 \\
\hfill \\
\, [\ion{O}{2}] $\lambda$3727 & =  & 0.62 (H$\alpha$ + [\ion{N}{2}] $\lambda\lambda$6548, 6583)  & CFRS-12$^3$  & $\sim83\%$ & 32  &  51.74 & 42.60 \\
\hfill \\ 
\, [\ion{O}{1}] $\lambda$6300 & =  & 0.08 H$\alpha$  & any        & $\sim92\%$ & 43  &  23.6 & 2.0 \\
\, [\ion{O}{1}] $\lambda$6300 & =  & 0.15 H$\alpha$  & 3$\sigma$  & $\sim77\%$ & 11  &  17.6 & 2.9 \\
\, [\ion{O}{1}] $\lambda$6300 & =  & 0.09 H$\alpha$  & 5$\sigma$  & $\sim88\%$ & 3  &  15.1 & 3.8 \\
\hfill \\
\tableline
\end{tabular}

\medskip
$^1$ {\footnotesize The correlation is measured with data which have
both EW(1) and EW(2) detected (see Fig.~\ref{plot2}).  In
the case of EW([\ion{O}{2}]) $= a $ EW(H$\alpha$ + [\ion{N}{2}]  $\lambda\lambda$6548, 6583), only
[\ion{O}{2}] and H$\alpha$ EW detection levels are considered;
[\ion{N}{2}]  $\lambda\lambda$6548, 6583 $=$ 1.33 [\ion{N}{2}]  $\lambda$6583 and [\ion{N}{2}]  $\lambda$6583 EW may be equal
to zero.}  \\ 
$^2$ {\footnotesize We excluded only Seyfert~1 from K92 
sample. If we exclude all AGN galaxies we find respectively $a = 0.41$,
K92, $\sim55$\%, 63, 60 and 25.} \\ 
$^3$ {\footnotesize The CFRS EWs are usually detected above 10 \AA.  
The correlation measured with the CFRS calibrated line fluxes gives also $a = 0.62$.}
\end{minipage}
\end{table*}

\begin{table*}
\begin{minipage}{150mm}
\caption{Mean and median of the [\ion{O}{2}] EW distributions (see Fig.~\ref{plot30})
 \label{table5}}
\begin{tabular}{lcc} 
Sample$^1$                 & [\ion{O}{2}] EW median  & [\ion{O}{2}]  EW mean \\ 
\tableline
\hfill \\  
SAPM [\ion{O}{2}]$>$0   & 9.6                & 12.6                  \\  
SAPM [\ion{O}{2}]$>$0, H$\alpha>0$ & 10.4    & 13.3                 \\  
SAPM [\ion{O}{2}]$>$0, H$\alpha\le0$ & 5.4   & 5.8                 \\  
\hfill \\ 
CFRS-12 [\ion{O}{2}]$>$0 & 29.4              &  34.9             \\
CFRS-12 [\ion{O}{2}]$>$0, H$\alpha>0$ & 42.6  & 41.7           \\
\hfill \\ 
CFRS-14 [\ion{O}{2}]$>$0 & 26.4             & 30.1          \\
\hfill \\
\tableline
\end{tabular}

\medskip
$^1$ Note that each sample has different properties, for instance the SAPM, blue selection, has a EW detection level $\sim 2$\AA,  $\langle z \rangle = 0.05 $ and $-24 <M(b_J) < -14$; the CFRS-12, red selection, has a EW detection level $\sim 10$\AA, $\langle z \rangle = 0.2 $ and $-21 <M(b_J) < -14$; and the CFRS-14, from red to blue selection, has a EW detection level $\sim 10$\AA, $\langle z \rangle = 0.6 $ and $-24 <M(b_J) < -14$. \\   
\end{minipage}
\end{table*}

\onecolumn

\noindent {\bf Figure 1.}{ Top panels: The complete rest-frame wavelength
range spanned by one SAPM spectrum at $z = 0.0534$. One can see the
small gaps in continuum at (observed frame) $\sim$[4360--4370]\AA, and
$\sim$[7000--7020]\AA\ (see Section 2.1).  Bottom-left panel: Zoomed
view of [\ion{O}{2}] $\lambda$3727. Bottom-right panel: Zoomed view of
[\ion{N}{2}] $\lambda$6548, H$\alpha$, [\ion{N}{2}] $\lambda$6583,
[\ion{S}{2}] $\lambda\lambda$6716, 6731 respectively. } \\

\noindent {\bf Figure 2.}{ Top panel: Log of the detection level of
EW(H$\alpha$), i.e. DL $=$ Log~(EW/$\sigma$(EW)), versus Log
EW(H$\alpha$). Lower panels: Same but for [\ion{O}{2}] $\lambda$3727,
for [\ion{S}{2}] $\lambda\lambda$6716, 6731, for [\ion{N}{2}]
$\lambda$6584, and for [\ion{O}{1}] $\lambda$6300. Vertical solid lines are the
respective EW medians (see Table~\ref{table4}).  The number of plotted
data and the 2, $3\sigma$ detection levels (dashed lines) are
indicated in each panel. } \\

\noindent {\bf Figure 3.}{ For the whole sample (1671), fractions of
the galaxy image covered by the 8\arcsec\ wide slit if the slit was
positioned along the major axis, $F(a)$, or the minor axis, $F(b)$, as
a function of the physical size at 25 mag arcsec$^{-2}$, $S_{25}$
(upper panels), the cosine inclination angle, cos($i$) (middle
panels), and the redshift, $z$ (lower panels).} \\

\noindent {\bf Figure 4.}{ Top panel: Physical size at 25 mag
arcsec$^{-2}$, $S_{25}$, as a function of the cosine inclination angle,
cos($i$). Middle panel: Distributions of cos($i$), the shaded
histogram is for unclassified galaxies (see Sect.~3.1). Bottom panel:
Fractions of ELG as a function cos($i$), the mean is $61\%$
(dashed-dotted line).} \\

\noindent {\bf Figure 5.}{ Log EW(H$\alpha$), and detection level of
EW(H$\alpha$) as a function of the fraction of the galaxy image
covered if the slit was positioned along the major axis, $F(a)$ (upper
panels), the cosine inclination angle, cos($i$) (middle panels), and
the redshift, $z$ (lower panels).} \\

\noindent {\bf Figure 6.}{ Top panel: N($b_J$) distributions, the shaded  
histogram is for the $82$ unclassified galaxies (see Sect.~3.1). 
Bottom panel: Fraction of ELG as a function of $b_J$, the
mean is $61\%$ (dashed-dotted line).} \\

\begin{figure}
\psfig{figure=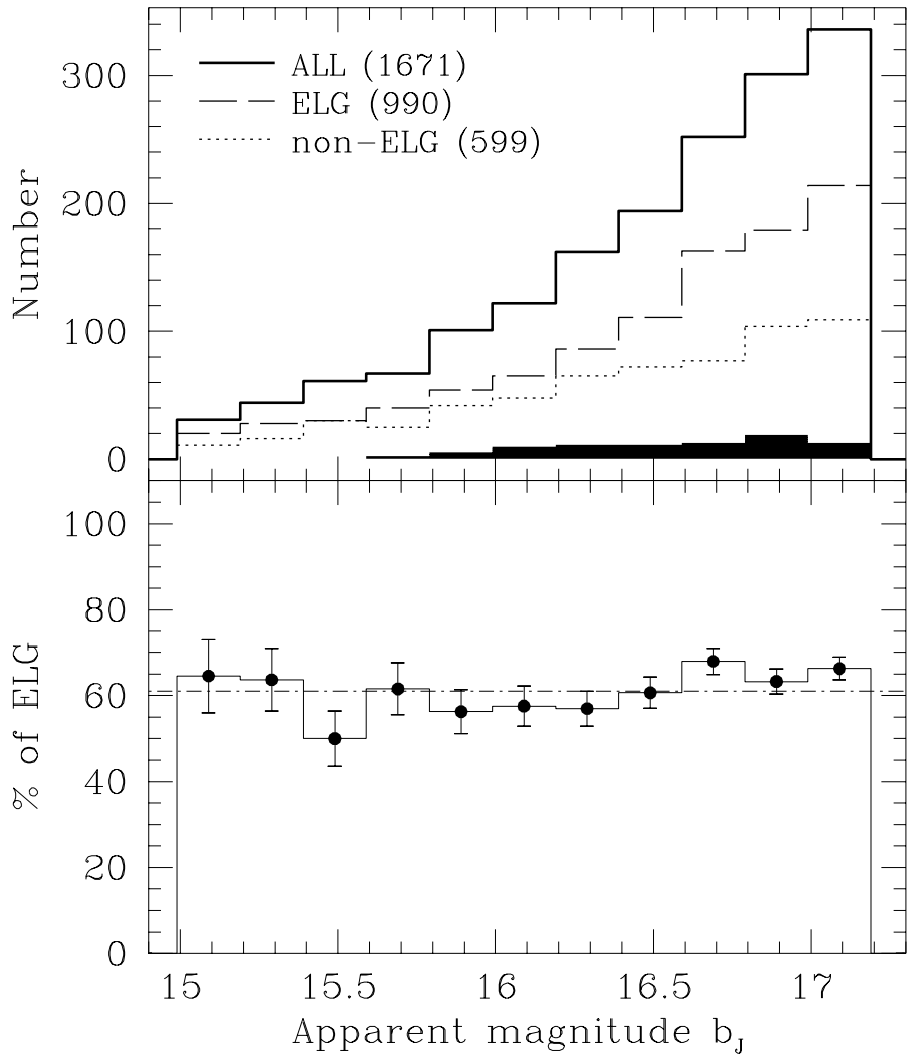,height=15cm} 
\caption[plot5]{ Top panel: N($b_J$) distributions, the shaded  
histogram is for the $82$ unclassified galaxies (see Sect.~3.1). 
Bottom panel: Fraction of ELG as a function of $b_J$, the
mean is $61\%$ (dashed-dotted line).
\label{plot5}}
\end{figure}

\begin{figure}
\psfig{figure=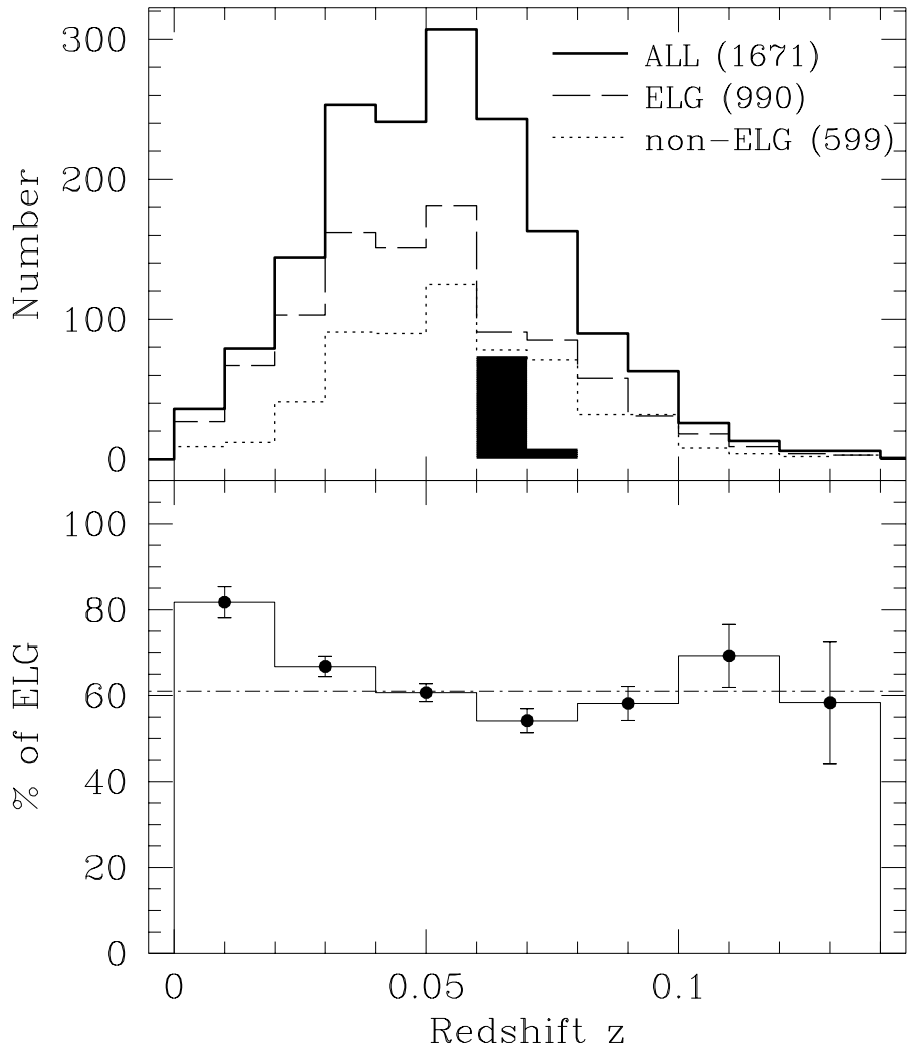,height=15cm} 
\caption[plot6]{ Top panel: N(z) distributions.  Bottom panel:
Fraction of ELG as a function of $z$. Notation the same as in
Figure~\ref{plot5}.
\label{plot6}}
\end{figure}

\begin{figure}
\psfig{figure=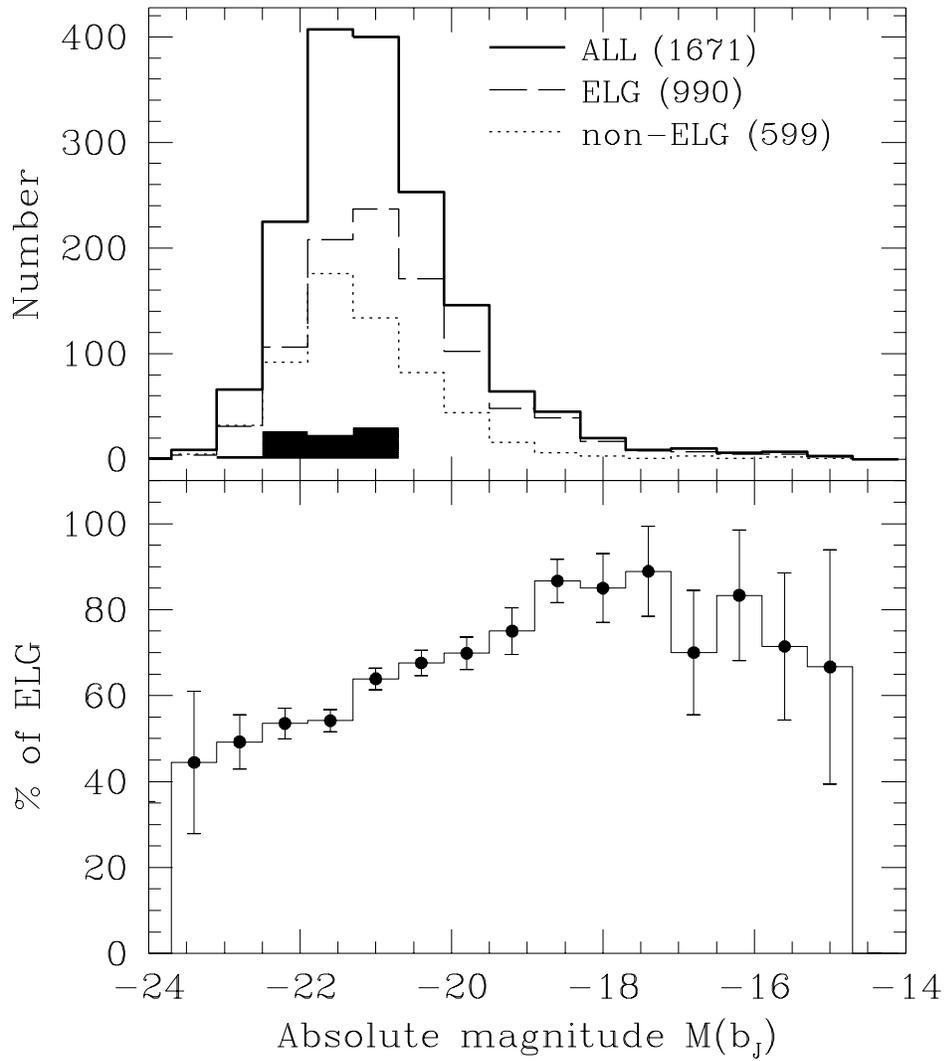,height=15cm} 
\caption[plot7]{ Top panel: N(M($b_J$)) distributions. Bottom panel:
Fraction of ELG as a function of M($b_J$).  Notation the same as in
Figure~\ref{plot5}.
\label{plot7}}
\end{figure}

\begin{figure}
\psfig{figure=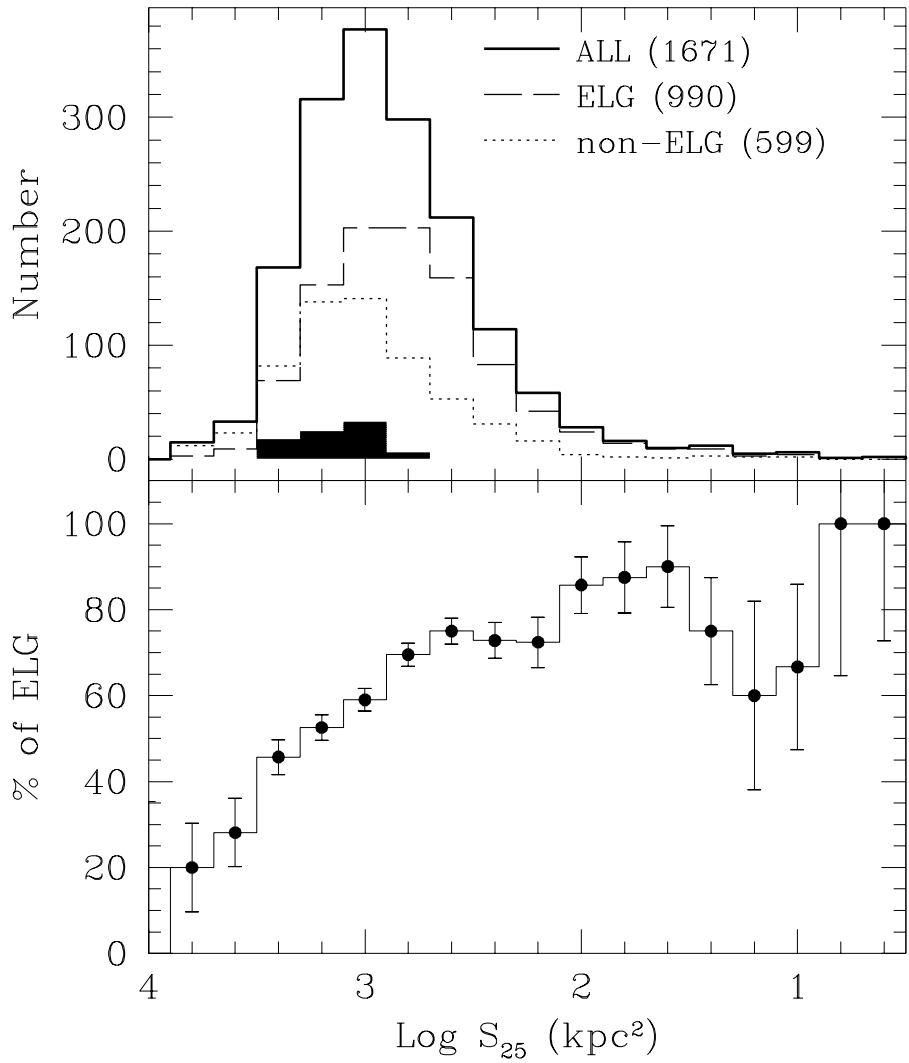,height=15cm} 
\caption[plot8]{ Top panel: Physical size $S_{25}$ distributions. 
Bottom panel: Fraction of ELG as a function of $S_{25}$.  Notation
the same as in Figure~\ref{plot5}.
\label{plot8}}
\end{figure}

\begin{figure}
\psfig{figure=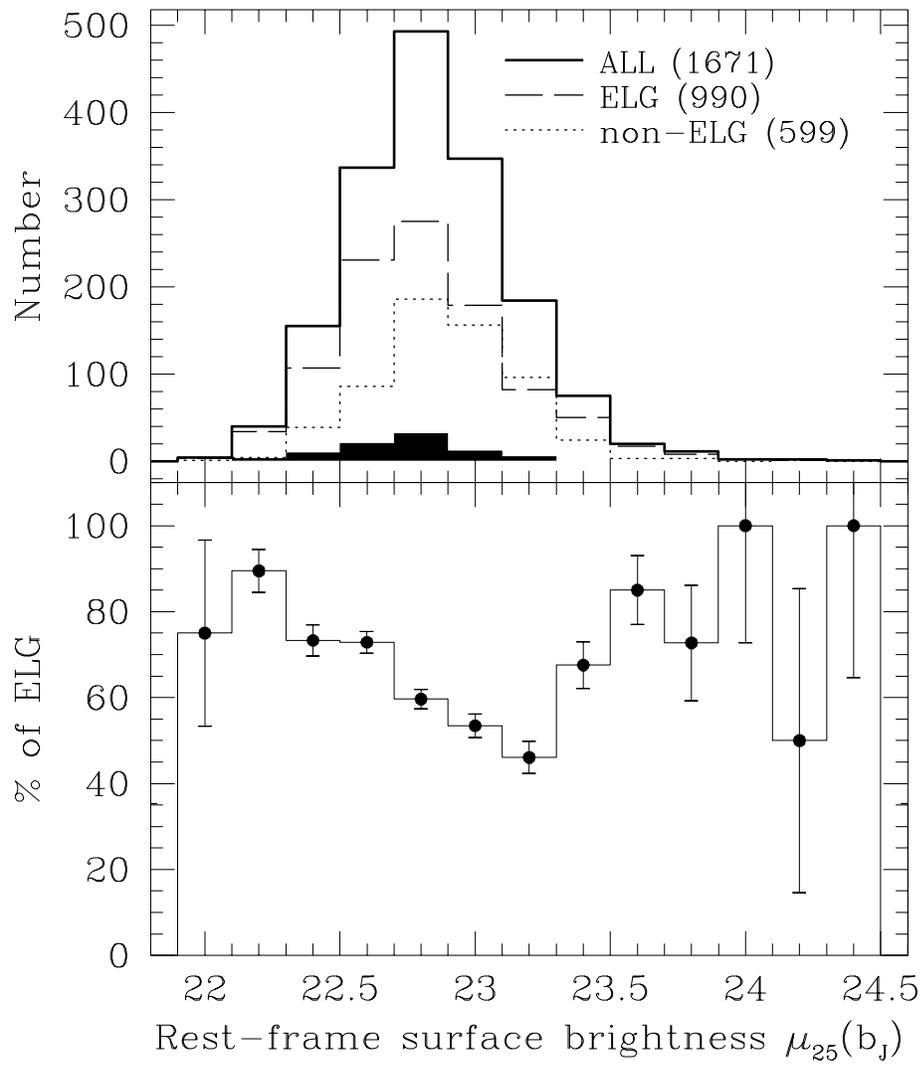,height=15cm} 
\caption[plot9]{ Top panel: Rest-frame surface brightness $\mu_{25}$
distributions.  Bottom panel: Fraction of ELG as a function of
$\mu_{25}$.  Notation the same as in Figure~\ref{plot5}.
\label{plot9}}
\end{figure}

\begin{figure}
\psfig{figure=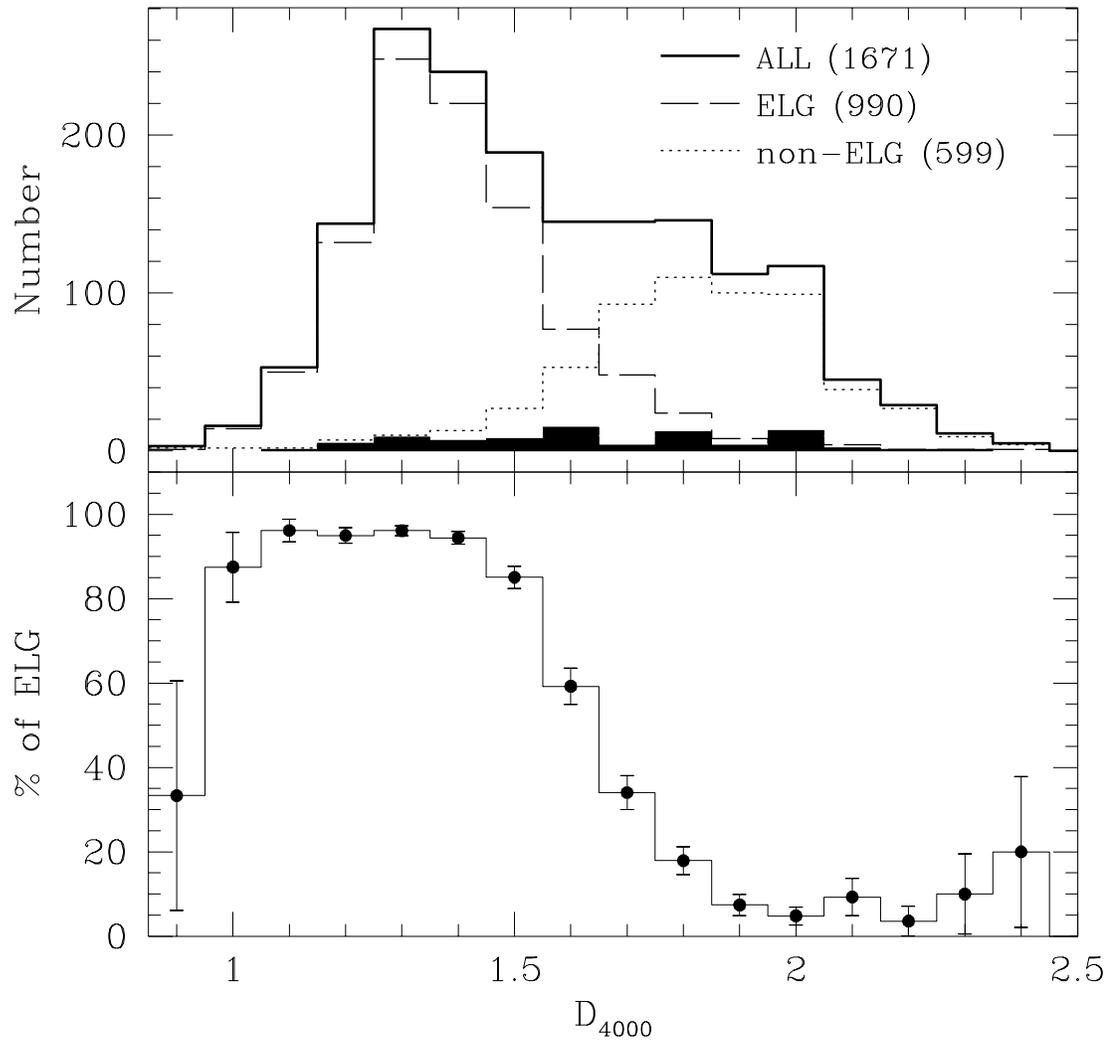,height=15cm}
\caption[plot4]{ Top panel: D$_{4000}$ distributions.  Bottom panel:
Fraction of ELG as a function of D$_{4000}$ .  Notation the same as in
Figure~\ref{plot5}.
\label{plot4}}
\end{figure}

\begin{figure}
\psfig{figure=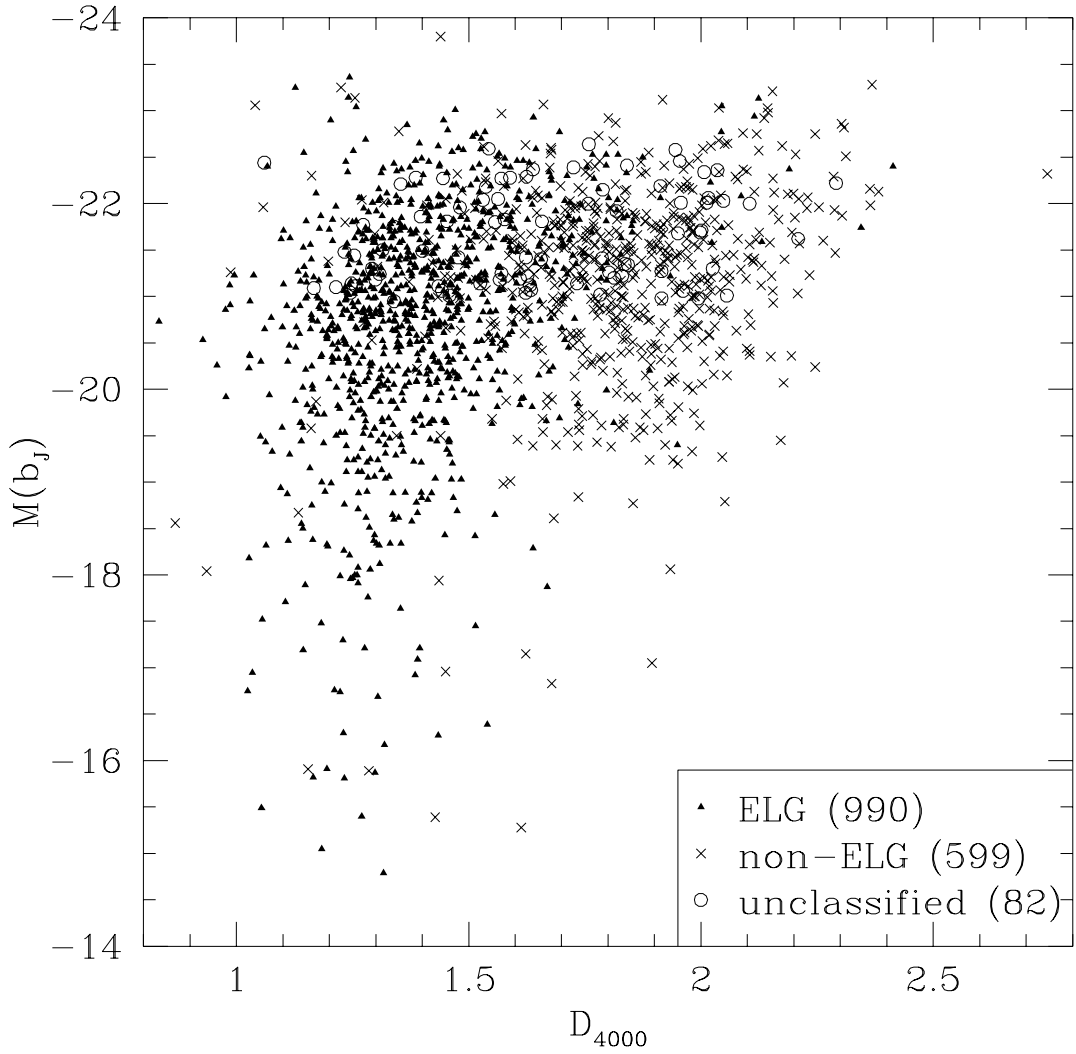,height=15cm}
\caption[plot4bis]{ M($b_J$) versus D$_{4000}$. 
\label{plot4bis}}
\end{figure}

\begin{figure}
\psfig{figure=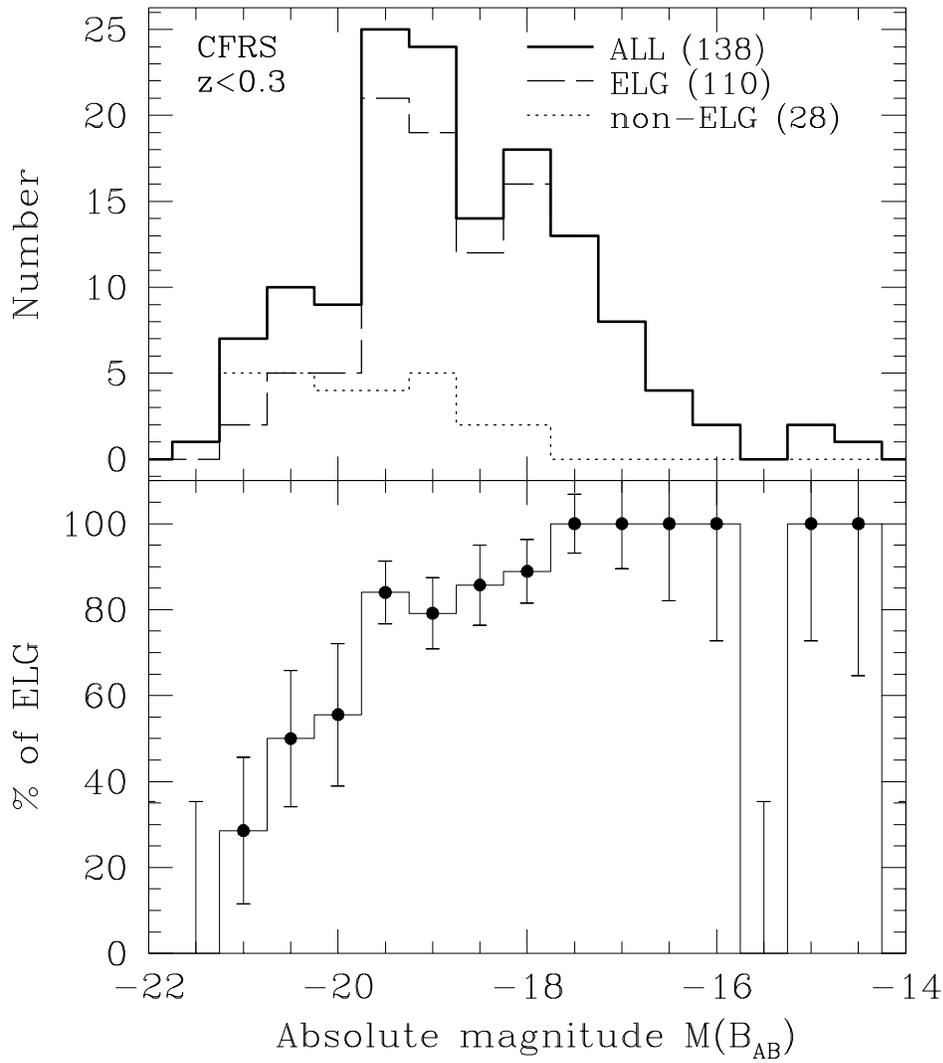,height=15cm} 
\caption[plot10]{ Top panel: N(M(B$_{AB}$)) distributions for the 138
CFRS galaxies at $z<0.3$, with H$\alpha$ detected (ELG) and with 
no H$\alpha$ detected (non-ELG). Bottom panel:
Fraction of CFRS ELG as a function of M(B$_{AB}$).
\label{plot10}}
\end{figure}

\begin{figure}
\psfig{figure=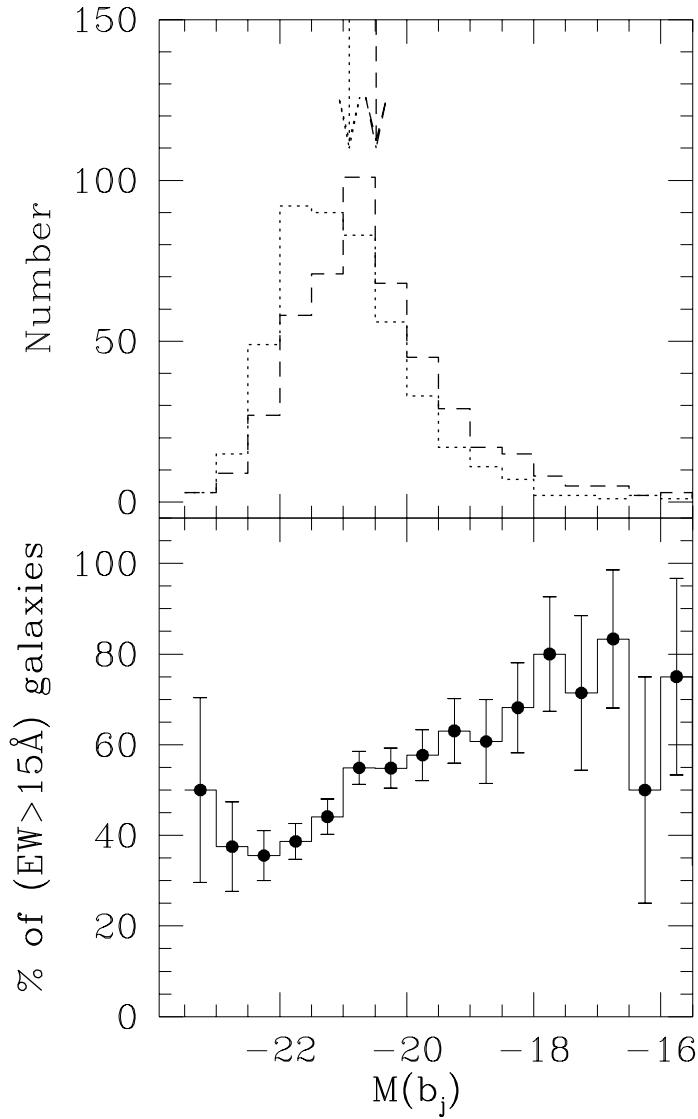,height=15cm} 
\caption[plot1213]{ Top panel: M($b_J$) distributions for the 934 ELG,
for the 467 EW(H$\alpha$) $>$ 15 \AA\ ELG (dashed line), and for the
467 EW(H$\alpha$) $\leq$ 15 \AA\ (dotted line).  Arrows are the
respective medians (see Table~\ref{table4}).  Bottom panel:
Fraction of EW(H$\alpha$) $>$ 15 \AA\ ELG as a function of M($b_J$).
\label{plot1213}}
\end{figure}

\begin{figure}
\psfig{figure=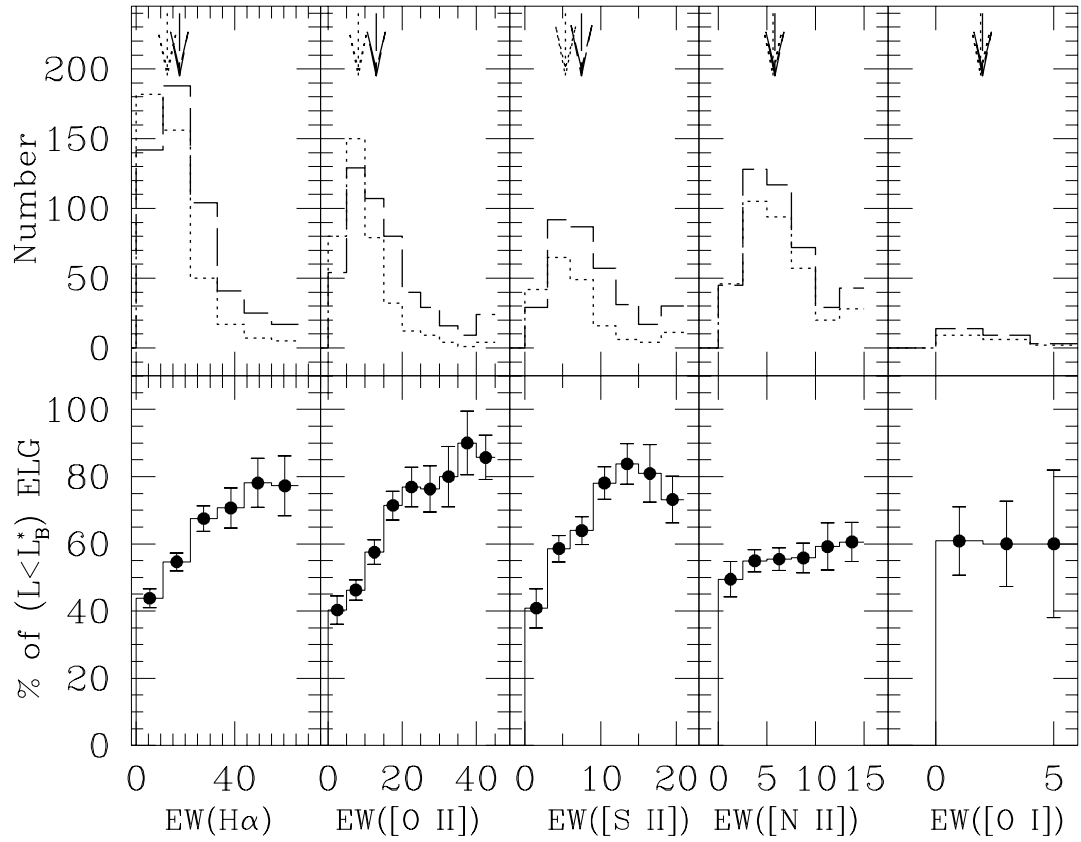,height=15cm}
\caption[plot142319]{ Top panels: EW distributions of H$\alpha$,
[\ion{O}{2}] $\lambda$3727, [\ion{S}{2}] $\lambda\lambda$6716, 6731,
[\ion{N}{2}] $\lambda$6583 and [{O}{1}] $\lambda$6300 for bright
$M(b_J) \leq -21$ (or $L\ge L^{*}_B$) ELG (dotted lines), and faint
($L < L^{*}_B$) ELG (dashed lines). Arrows are the respective medians
(see Table~\ref{table4}). Bottom panels: Respective fractions of faint
ELG as a function of EW.
\label{plot142319}}
\end{figure}

\begin{figure}
\psfig{figure=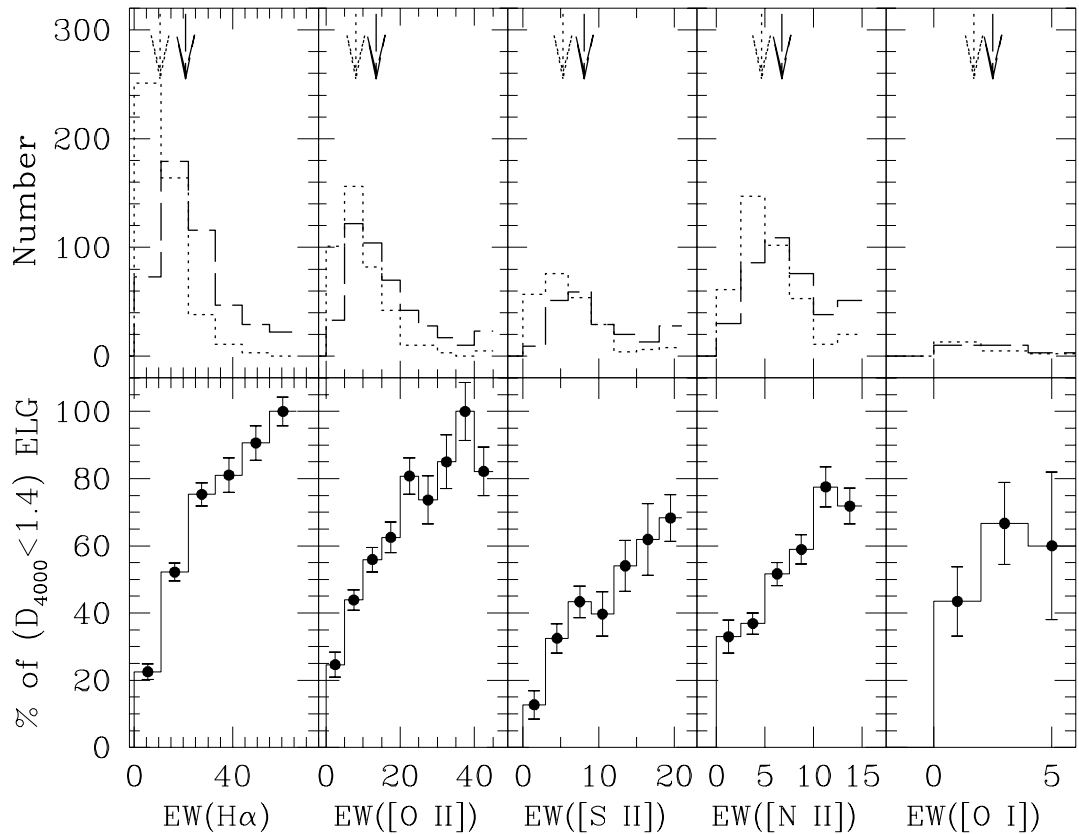,height=15cm} 
\caption[break6]{ Top panels: EW distributions of H$\alpha$,
[\ion{O}{2}] $\lambda$3727, [\ion{S}{2}] $\lambda\lambda$6716, 6731,
[\ion{N}{2}] $\lambda$6583 and [\ion{O}{1}] $\lambda6300$ for high
$D_{4000} \ge 1.4$ ELG (dotted lines), and low $D_{4000} < 1.4$ ELG
(dashed lines). Arrows are the respective medians (see
Table~\ref{table4}). Bottom panels: Respective fractions of low
D$_{4000}$ ELG as a function of EW.
\label{break6}}
\end{figure}

\begin{figure}
\psfig{figure=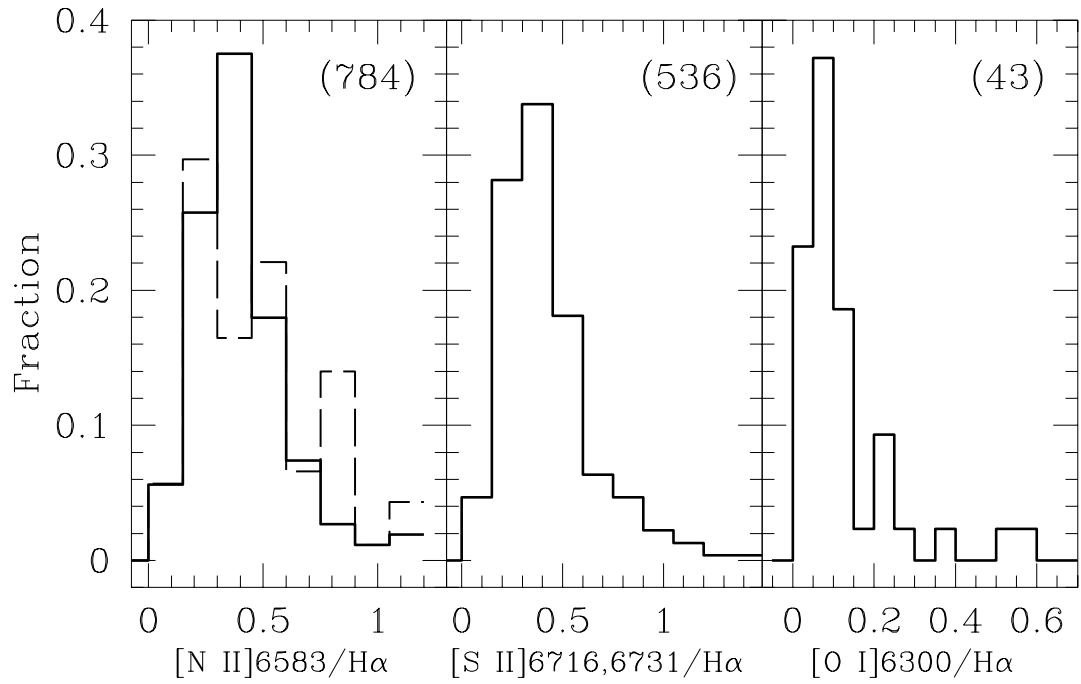,height=15cm} 
\caption[plot15b]{ Distribution of the EW ratios for [\ion{N}{2}]
$\lambda$6583/H$\alpha$, [\ion{S}{2}] $\lambda\lambda$6716,6731/H$\alpha$ 
and [\ion{O}{1}] $\lambda$6300/H$\alpha$ (solid line). 
The dashed line is for 57 galaxies from K92's sample weighted as
described in Section 6.1.  
\label{plot15b}}
\end{figure}

\begin{figure}
\psfig{figure=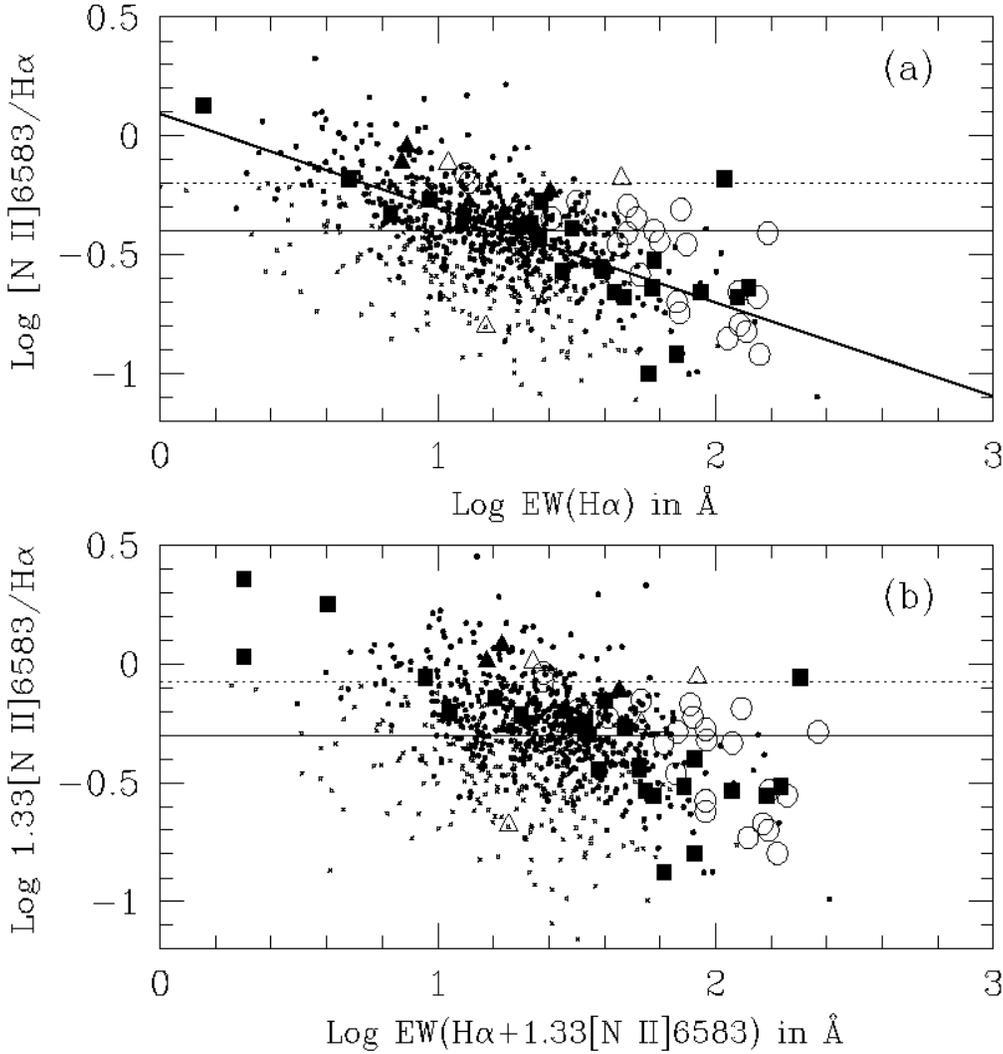,height=15cm}
\caption[plot18new]{ (a) Log [\ion{N}{2}] $\lambda$6583/H$\alpha$
versus Log EW(H$\alpha$).  (b) Log 1.33 [\ion{N}{2}] $\lambda$6583/H$\alpha$
versus Log EW(H$\alpha$ + 1.33[\ion{N}{2}] $\lambda$6583).  Dots are SAPM ($>
3\sigma$) data, crosses are SAPM ($< 3\sigma$) data.  Open symbols are
K92 high-resolution data, filled symbols are K92 low-resolution
data. Triangles are AGN and circles are normal galaxies in K92 sample.
The dotted line is the empirical separation between AGN and
\hbox{H\,{\sc ii}} galaxies. The solid line is the commonly used
average value for [\ion{N}{2}]/H$\alpha$.  The diagonal line is
our best fit to the correlation.
\label{plot18new}}
\end{figure}

\noindent {\bf Figure 7.}{ Top panel: N(z) distributions.  Bottom
panel: Fraction of ELG as a function of $z$. Notation the same as in
Figure~\ref{plot5}.} \\

\noindent {\bf Figure 8.}{ Top panel: N(M($b_J$))
distributions. Bottom panel: Fraction of ELG as a function of
M($b_J$).  Notation the same as in Figure~\ref{plot5}.} \\

\noindent {\bf Figure 9.}{ Top panel: Physical size $S_{25}$
distributions. Bottom panel: Fraction of ELG as a function of
$S_{25}$.  Notation the same as in Figure~\ref{plot5}.} \\

\noindent {\bf Figure 10.}{ Top panel: Rest-frame surface brightness
$\mu_{25}$ distributions.  Bottom panel: Fraction of ELG as a function
of $\mu_{25}$.  Notation the same as in Figure~\ref{plot5}.} \\

\noindent {\bf Figure 11.}{ Top panel: D$_{4000}$ distributions.
Bottom panel: Fraction of ELG as a function of D$_{4000}$ .  Notation
the same as in Figure~\ref{plot5}.} \\

\noindent {\bf Figure 12.}{ M($b_J$) versus D$_{4000}$} \\

\noindent {\bf Figure 13.}{ Top panel: N(M(B$_{AB}$)) distributions
for the 138 CFRS galaxies at $z<0.3$, with H$\alpha$ detected (ELG)
and with no H$\alpha$ detected (non-ELG). Bottom panel: Fraction of
CFRS ELG as a function of M(B$_{AB}$). } \\

\noindent {\bf Figure 14.}{ Top panel: M($b_J$) distributions for the
934 ELG, for the 467 EW(H$\alpha$) $>$ 15 \AA\ ELG (dashed line), and
for the 467 EW(H$\alpha$) $\leq$ 15 \AA\ (dotted line).  Arrows are
the respective medians (see Table~\ref{table4}).  Bottom panel:
Fraction of EW(H$\alpha$) $>$ 15 \AA\ ELG as a function of M($b_J$).} \\

\noindent {\bf Figure 15.}{ Top panels: EW distributions of H$\alpha$, 
[\ion{O}{2}] $\lambda$3727, [\ion{S}{2}] $\lambda\lambda$6716, 6731,
[\ion{N}{2}] $\lambda$6583 and [\ion{O}{1}] $\lambda$6300 for bright
$M(b_J) \leq -21$ (or $L \ge L^{*}_B$) ELG (dotted lines), and faint
($L < L^{*}_B$) ELG (dashed lines). Arrows are the respective medians
(see Table~\ref{table4}). Bottom panels: Respective fractions of faint
ELG as a function of EW.} \\

\noindent {\bf Figure 16.}{ Top panels: EW distributions of H$\alpha$,
[\ion{O}{2}] $\lambda$3727, [\ion{S}{2}] $\lambda\lambda$6716, 6731
and [\ion{N}{2}] $\lambda$6583 lines for high $D_{4000} \ge 1.4$ ELG
(dotted lines), and low $D_{4000} < 1.4$ ELG (dashed lines). Arrows
are the respective medians (see Table~\ref{table4}). Bottom panels:
Respective fractions of low D$_{4000}$ ELG as a function of EW. } \\

\noindent {\bf Figure 17.}{ Distribution of the EW ratios for
[\ion{N}{2}] $\lambda$6583/H$\alpha$, [\ion{S}{2}]
$\lambda\lambda$6716,6731/H$\alpha$ and [\ion{O}{1}]
$\lambda$6300/H$\alpha$ (solid line).  The dashed line is for 57
galaxies from K92's sample weighted as described in Section 6.1.} \\

\noindent {\bf Figure 18.}{ (a) Log [\ion{N}{2}]
$\lambda$6583/H$\alpha$ versus Log EW(H$\alpha$).  (b) Log 1.33
[\ion{N}{2}] $\lambda$6583/H$\alpha$ versus Log EW(H$\alpha$) +
1.33EW([\ion{N}{2}] $\lambda$6583).  Dots are SAPM ($> 3\sigma$) data,
crosses are SAPM ($< 3\sigma$) data.  Open symbols are K92
high-resolution data, filled symbols are K92 low-resolution
data. Triangles are AGN and circles are normal galaxies in K92 sample.
The dotted line is the empirical separation between AGN and
\hbox{H\,{\sc ii}} galaxies. The solid line is the commonly used
average value for [\ion{N}{2}]/H$\alpha$.  The diagonal line is
our best fit to the correlation.} \\

\noindent {\bf Figure 19.}{ Log EW([\ion{N}{2}] $\lambda$6583) versus
Log EW(H$\alpha$).  Small symbols are data detected below $3\sigma$,
large symbols are $> 3\sigma$ data. AGN candidates ([\ion{N}{2}]
$\lambda$6563/H$\alpha > 0.63$) are denoted by stars and non-AGN by
filled circles. The solid line is the correlation using all ($>
3\sigma$) data; EW([\ion{N}{2}]) $\approx 0.4$ EW(H$\alpha$) (see
Table~\ref{table3}). The average $1\sigma$ EW errorbars are also
shown.} \\

\begin{figure}
\psfig{figure=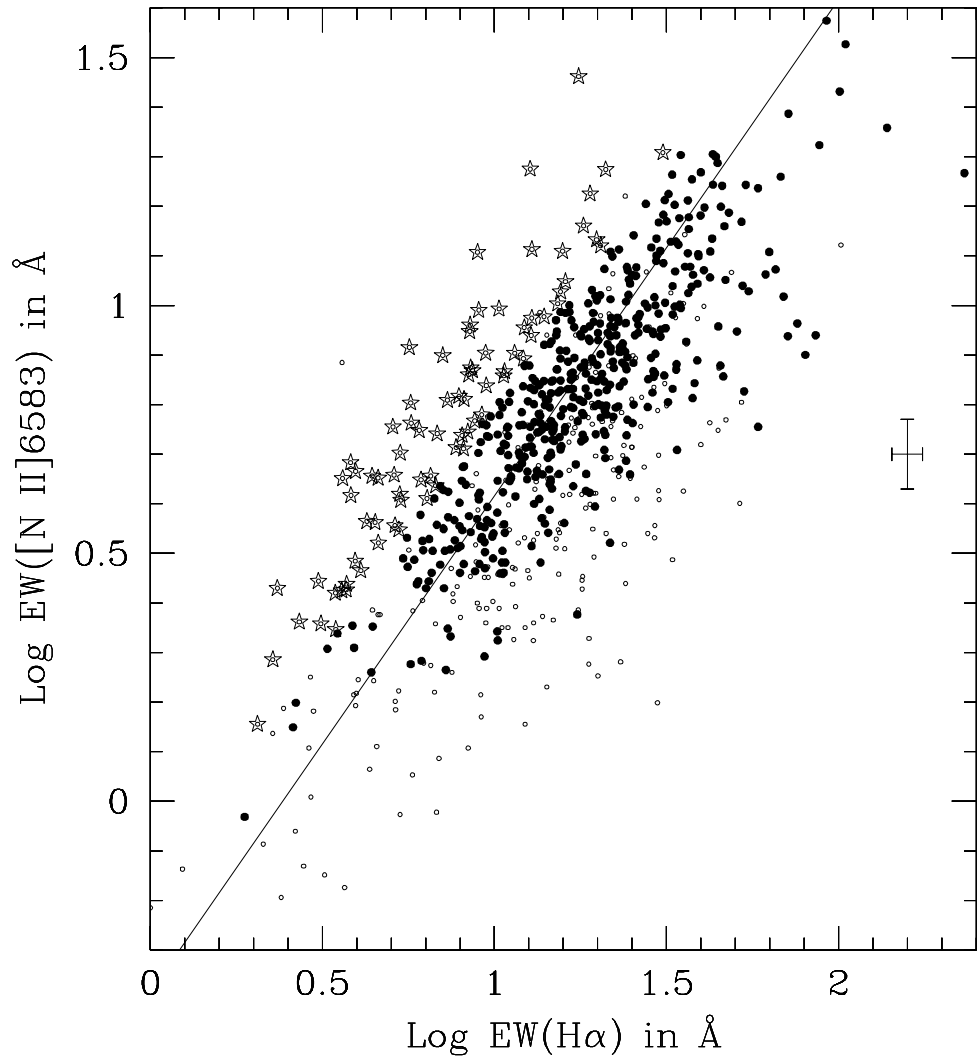,height=15cm} 
\caption[plot25]{ Log EW([\ion{N}{2}] $\lambda$6583) versus Log EW(H$\alpha$).
Small symbols are data detected below $3\sigma$, large symbols are $>
3\sigma$ data. AGN candidates ([\ion{N}{2}] $\lambda$6563/H$\alpha > 0.63$) are
denoted by stars and non-AGN by filled circles. The solid line is the
correlation using all ($> 3\sigma$) data; EW([\ion{N}{2}])
$\approx 0.4$ EW(H$\alpha$) (see Table~\ref{table3}). The average
$1\sigma$ EW errorbars are also shown.
\label{plot25}}
\end{figure}

\begin{figure}
\psfig{figure=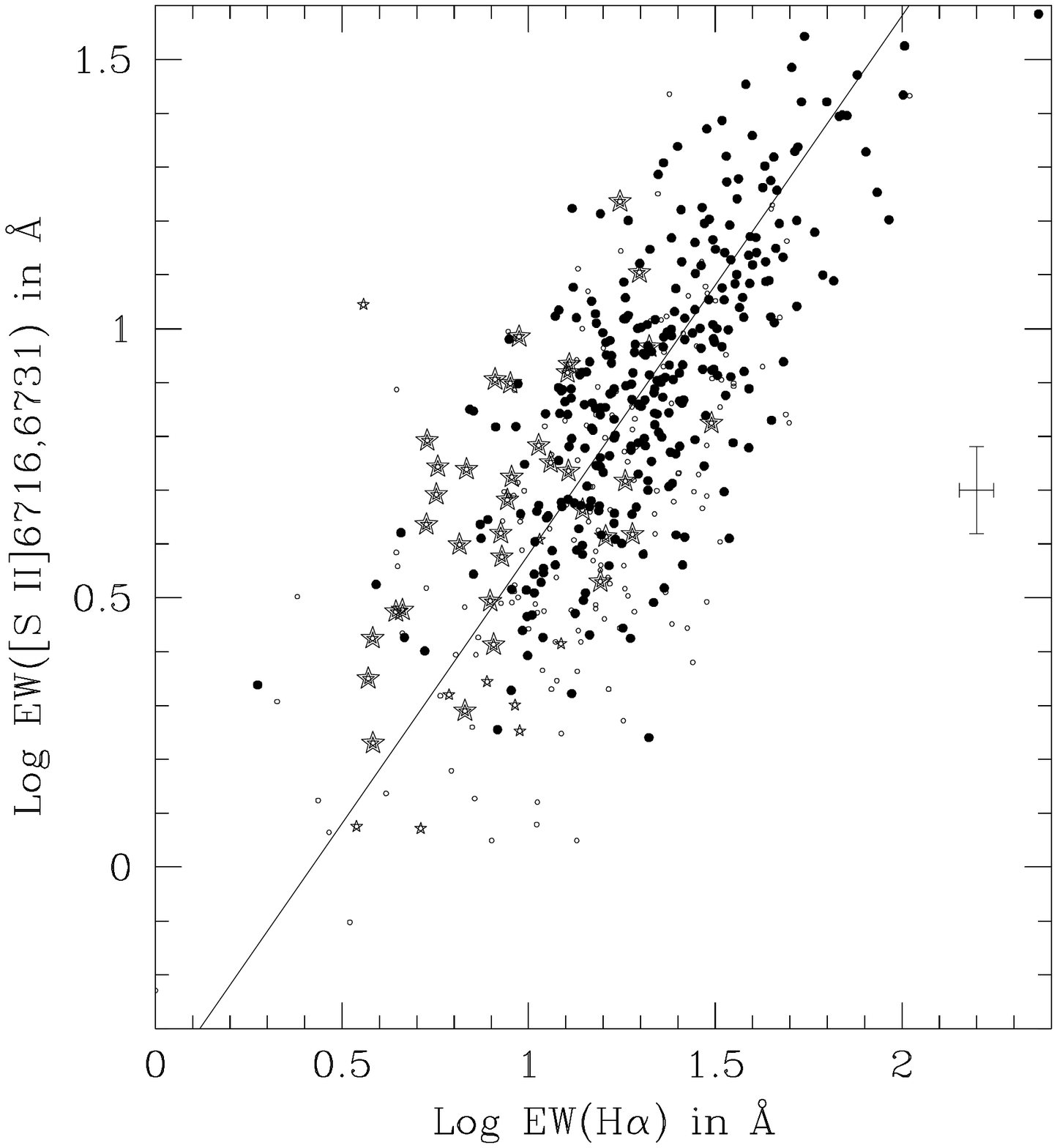,height=15cm} 
\caption[plot25a]{ Log EW([\ion{S}{2}] $\lambda\lambda$6716, 6731)
versus Log EW(H$\alpha$).  Same notation as
Figure~\ref{plot25}. EW([\ion{S}{2}]) $\approx 0.4$ EW(H$\alpha$) (see
Table~\ref{table3}).
\label{plot25a}}
\end{figure}

\begin{figure}
\psfig{figure=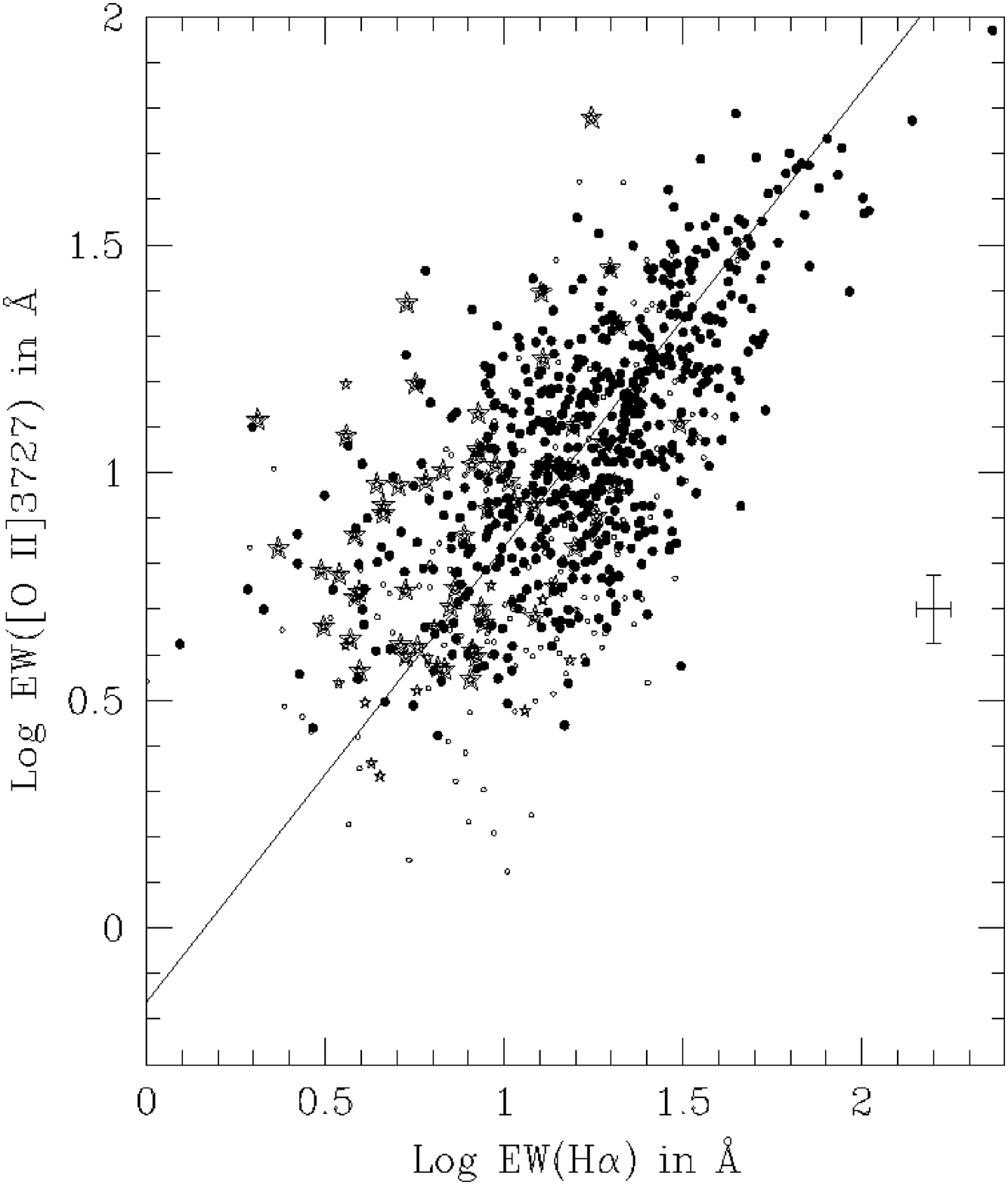,height=15cm} 
\caption[plot25b]{ Log EW([\ion{O}{2} $\lambda$3727]) versus Log EW(H$\alpha$).
Same notation as Figure~\ref{plot25}.  EW([\ion{O}{2}]) $\approx 0.7$
EW(H$\alpha$) (see Table~\ref{table3}).
\label{plot25b}}
\end{figure}

\noindent {\bf Figure 20.}{ Log EW([\ion{S}{2}] $\lambda\lambda$6716,
6731) versus Log EW(H$\alpha$).  Same notation as
Figure~\ref{plot25}. EW([\ion{S}{2}]) $\approx 0.4$
EW(H$\alpha$) (see Table~\ref{table3}).}  \\

\noindent {\bf Figure 21.}{ Log EW([\ion{O}{2}] $\lambda$3727) versus Log
EW(H$\alpha$).  Same notation as Figure~\ref{plot25}.
EW([\ion{O}{2}]) $\approx 0.7$ EW(H$\alpha$) (see
Table~\ref{table3}).} \\

\noindent {\bf Figure 22.}{ Log EW([\ion{O}{2}] $\lambda$3727) versus Log
EW(H$\alpha$ + [\ion{N}{2}] $\lambda\lambda$6548, 6583).  Same notation as
Figure~\ref{plot25}.  EW([\ion{O}{2}]) $\approx 0.5$ EW(H$\alpha$ +
[\ion{N}{2}]) (see Table~\ref{table3}).} \\

\noindent {\bf Figure 23.}{ Log EW([\ion{O}{1}] $\lambda6300$) versus Log
EW(H$\alpha$).  Same notation as Figure~\ref{plot25}.
EW([\ion{O}{1}]) $\approx 0.1$ EW(H$\alpha$); note the correlation 
is very poor (see Table~\ref{table3}).} \\

\begin{figure}
\psfig{figure=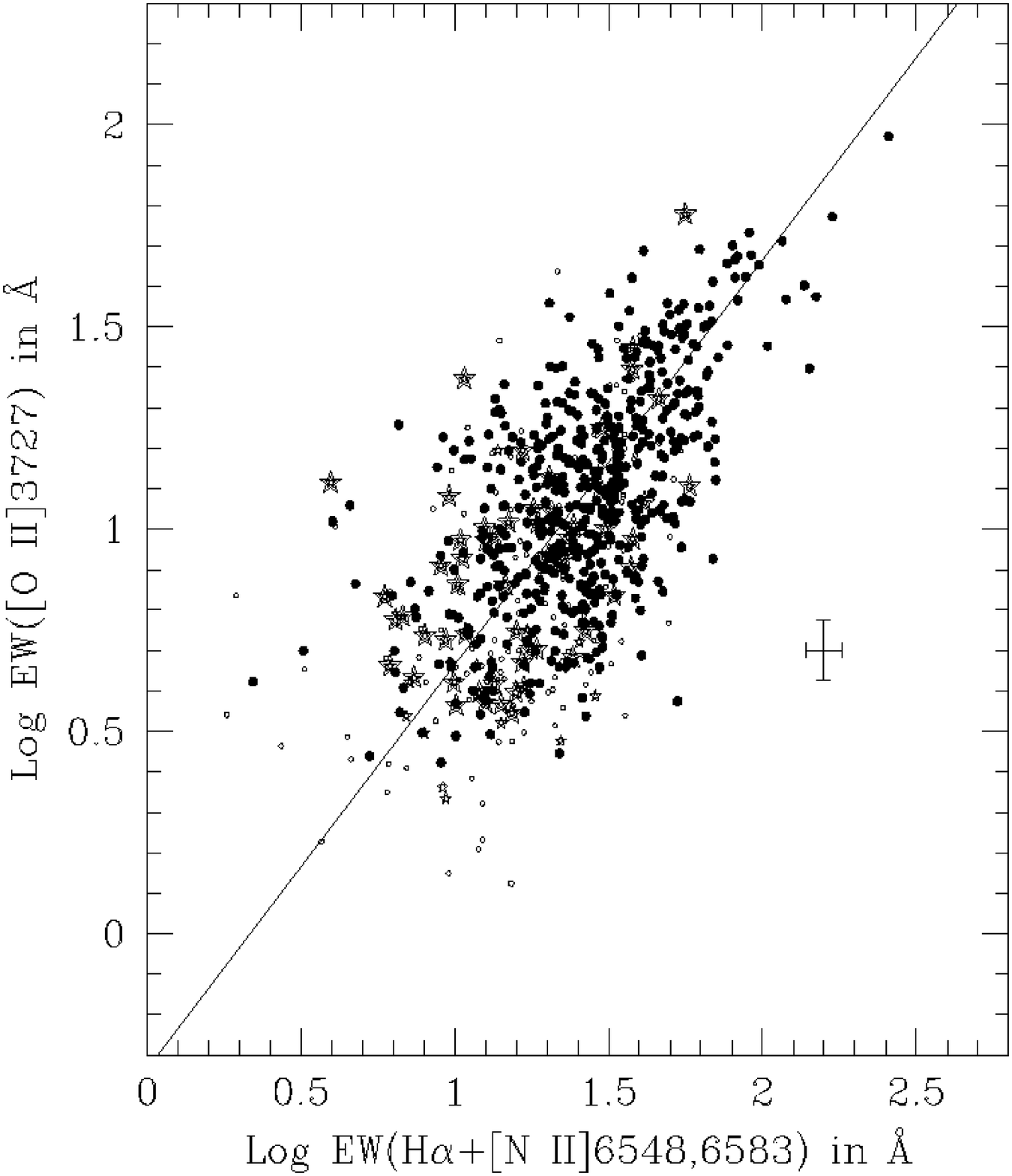,height=15cm} 
\caption[plot25c]{ Log EW([\ion{O}{2}] $\lambda3727$) versus Log
EW(H$\alpha$ + [\ion{N}{2}] $\lambda\lambda$6548, 6583).  Same
notation as Figure~\ref{plot25}.  EW([\ion{O}{2}]) $\approx 0.5$
EW(H$\alpha$ + [\ion{N}{2}]) (see Table~\ref{table3}).
\label{plot25c}}
\end{figure}

\begin{figure}
\psfig{figure=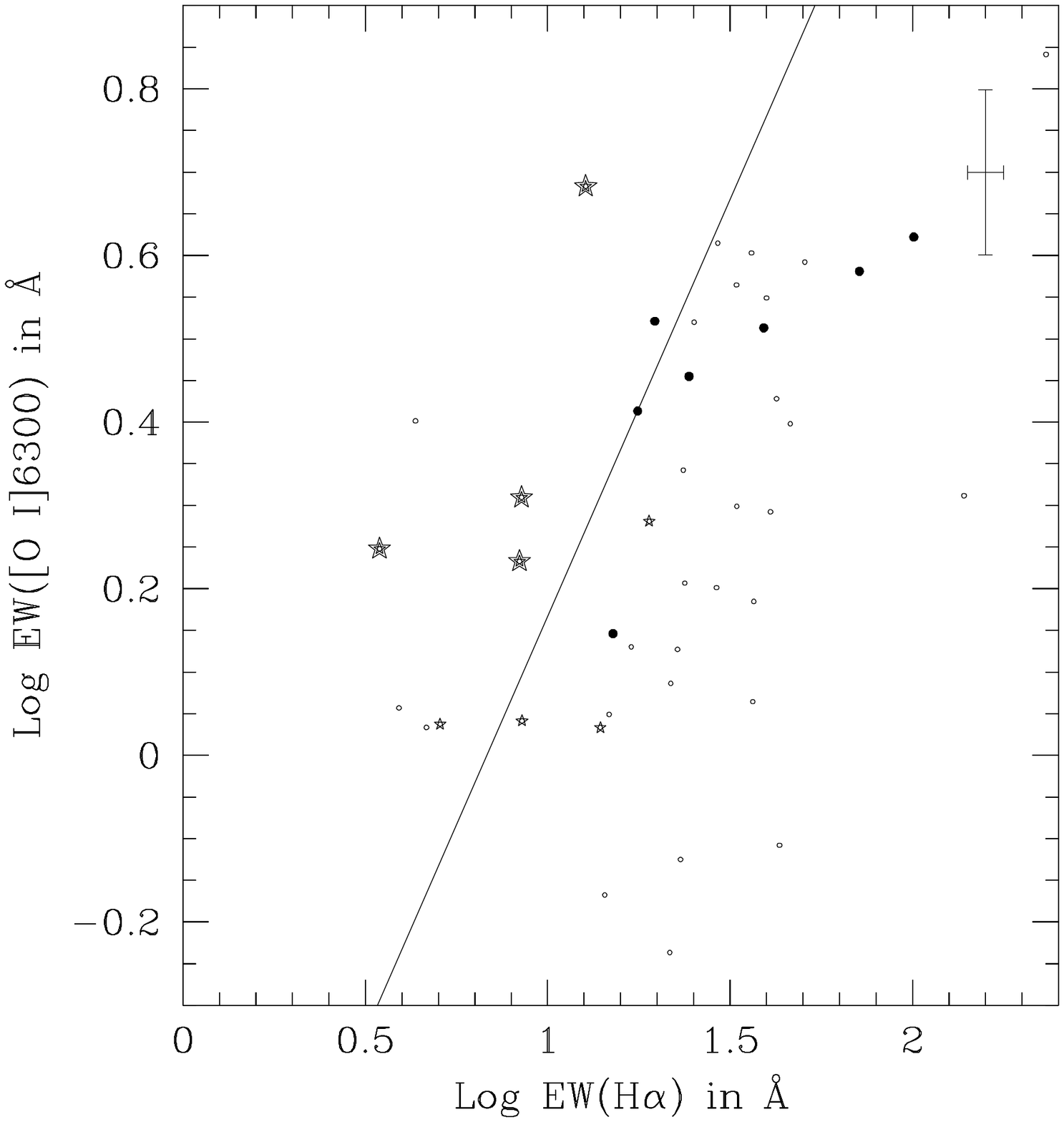,height=15cm} 
\caption[plot25e]{ Log EW([\ion{O}{1}] $\lambda6300$) versus Log
EW(H$\alpha$).  Same notation as Figure~\ref{plot25}.
EW([\ion{O}{1}]) $\approx 0.1$ EW(H$\alpha$); note the correlation 
is very poor (see Table~\ref{table3}).
\label{plot25e}}
\end{figure}

\begin{figure}
\psfig{figure=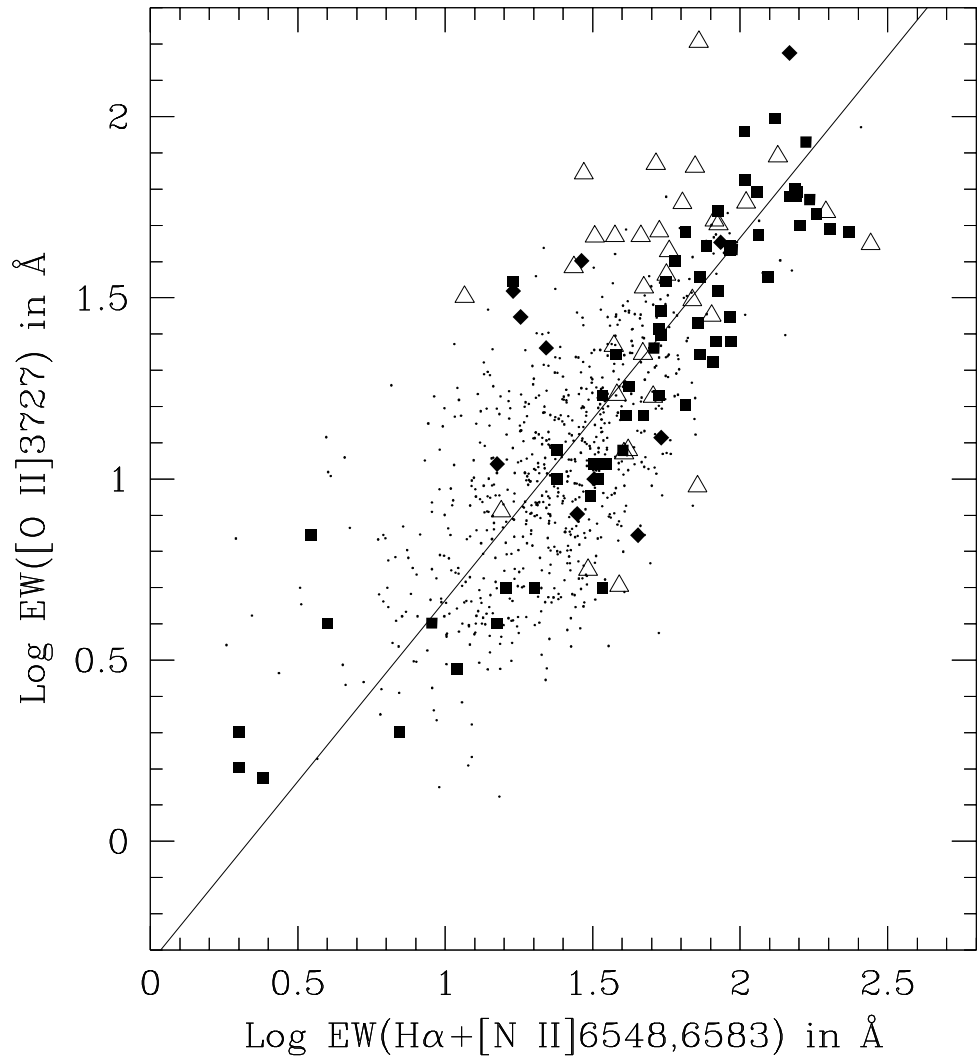,height=15cm} 
\caption[plot25d]{ Log EW([\ion{O}{2}] $\lambda3727$) versus Log
EW(H$\alpha$ + [\ion{N}{2}] $\lambda\lambda$6548, 6583). Dots are SAPM
data, and the line shows the correlation from Figure~\ref{plot25c}.
Filled symbols show K92 data (diamonds are AGN galaxies excluding
Seyfert1) and open symbols show CFRS-12 data.
\label{plot25d}}
\end{figure}

\begin{figure}
\psfig{figure=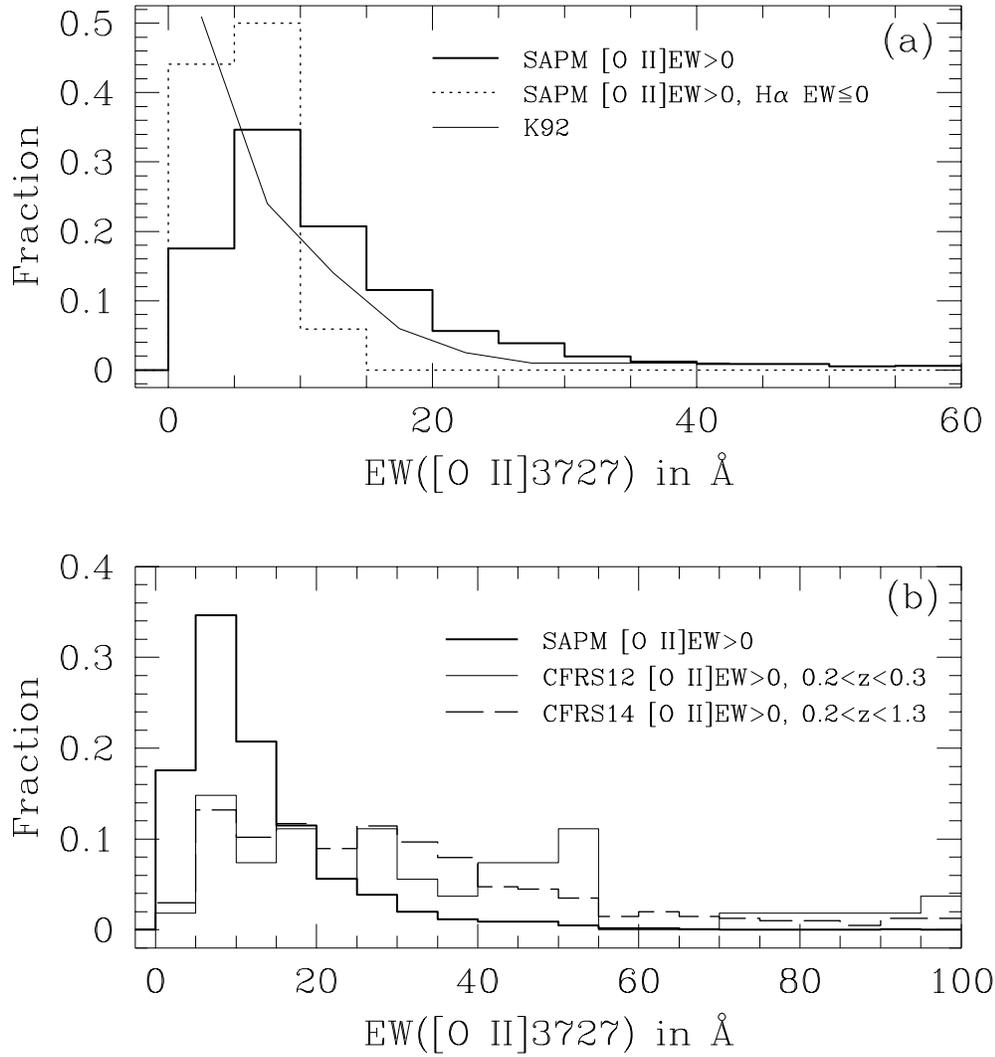,height=15cm} 
\caption[plot30]{ Relative distributions of EW([\ion{O}{2}]
$\lambda3727$). (a) For the 1008 SAPM spectra with [\ion{O}{2}]
detected (solid line), and for the 68 spectra with [\ion{O}{2}] but
not H$\alpha$ detected (dotted line). The curve is reproduced from
figure 9 in Kennicutt 1992, which shows the distribution of very
nearby galaxies. (b) For the 1008 SAPM data (thick-solid line), for 55
CFRS-12 data at $0.2<z < 0.3$ (thin-solid line), and for 403 CFRS-14
data ($0.2 < z < 1.3$; Hammer et al. 1997) (dashed line). Medians of
the distributions are listed in Table~\ref{table5}.
\label{plot30}}
\end{figure}

\noindent {\bf Figure 24.}{ Log EW([\ion{O}{2}] $\lambda$3727) versus
Log EW(H$\alpha$ + [\ion{N}{2}] $\lambda\lambda$6548, 6583). Dots are
SAPM data, and the line shows the correlation from
Figure~\ref{plot25c}.  Filled symbols show K92 data (diamonds are AGN
galaxies excluding Seyfert1) and open symbols show CFRS-12 data.} \\

\noindent {\bf Figure 25.}{ Relative distributions of EW([\ion{O}{2}]
$\lambda3727$). (a) For the 1008 SAPM spectra with [\ion{O}{2}]
detected (solid line), and for the 68 spectra with [\ion{O}{2}] but
not H$\alpha$ detected (dotted line). The curve is reproduced from
figure 9 in Kennicutt 1992, which shows the distribution of very
nearby galaxies. (b) For the 1008 SAPM data (thick-solid line), for 55
CFRS-12 data at $0.2<z < 0.3$ (thin-solid line), and for 403 CFRS-14
data ($0.2 < z < 1.3$; Hammer et al. 1997) (dashed line). Medians of
the distributions are listed in Table~\ref{table5}.} \\

\label{lastpage}

\onecolumn

\clearpage
%fig 1 
\clearpage
%fig 2
\clearpage
%fig 3
\clearpage
%fig 4
\clearpage
%fig 5
\clearpage
%fig 6 
\clearpage
%fig 7 
\clearpage
%fig 8
\clearpage
%fig 9 
\clearpage
%fig 10
\clearpage
%fig 11 
\clearpage
%fig 12 
\clearpage
%fig 13
\clearpage
%fig 14
\clearpage
%fig 15
\clearpage
%fig 16 
\clearpage
%fig 17
\clearpage
%fig 18 
\clearpage
%fig 19
\clearpage
%fig 20
\clearpage
%fig 21
\clearpage
%fig 22
\clearpage
%fig 23
\clearpage
%fig 24
\clearpage
%fig 25

\end{document}